\documentclass[twocolumn]{aastex62}
\usepackage{url}
\usepackage[nolist]{acronym}
\usepackage{appendix}
\usepackage{amsmath}

\graphicspath{{./figures_revision/}{}}

\shorttitle{STEREO/SEPT electron spectra}
\shortauthors{Dresing et al.}

\begin{document}
\begin{acronym}
	\acro{ar}[AR]{Active Region}
	\acroplural{ar}[ARs]{Active Regions}
	\acro{cir}[CIR]{Corotating Interaction Region}
	\acroplural{cir}[CIRs]{Corotating Interaction Regions}
	\acro{cme}[CME]{Coronal Mass Ejection}
	\acroplural{cme}[CMEs]{Coronal Mass Ejections}
	\acro{icme}[ICME]{Interplanetary Coronal Mass Ejection}
	\acro{gcs}[GCS]{Graduated Cylindrical Shell}
	\acro{gle}[GLE]{Ground Level Enhancement}
	\acroplural{gle}[GLEs]{Ground Level Enhancements}
	\acro{het}[HET]{High Energy Telescope}
	\acro{hxr}[HXR]{Hard X-Ray}
	\acro{ip}[IP]{interplanetary}
	\acroplural{icme}[ICMEs]{Interplanetary Coronal Mass Ejections}
	\acro{mhd}[MHD]{magnetohydrodynamic}
	\acro{nr}[NR]{near-relativistic}
	\acro{sep}[SEP]{Solar Energetic Particle}
	\acroplural{sep}[SEPs]{Solar Energetic Particles}
	\acro{sept}[SEPT]{Solar Electron and Proton Telescope}
	\acro{sir}[SIR]{Stream Interaction Region}
	\acroplural{sir}[SIRs]{Stream Interaction Regions}
	\acro{sta}[STA]{STEREO~A}
	\acro{stb}[STB]{STEREO~B}
\end{acronym}

\newcommand{\nall}{781}  
\newcommand{\nspl}{344}  
\newcommand{\mspl}{-3.47} 
\newcommand{\msplWid}{1.15}   
\newcommand{\ndpl}{437}  
\newcommand{\mLow}{-2.53} 
\newcommand{\mLowWid}{0.87} 
\newcommand{\mHigh}{-3.93} 
\newcommand{\mHighWid}{1.53} 
\newcommand{\mbreak}{126} 
\newcommand{\mbreakWid}{56}  
\newcommand{\nDown}{781} 
\newcommand{\mDown}{-2.94} 
\newcommand{\mDownWid}{1.19} 
\newcommand{\nUp}{420} 
\newcommand{\mUp}{-3.55} 
\newcommand{\mUpWid}{1.45} 

\newcommand{\nallImp}{228}  
\newcommand{\nsplImp}{132}  
\newcommand{\msplImp}{-3.94} 
\newcommand{\msplImpWid}{1.42}
\newcommand{\ndplImp}{96}  
\newcommand{\mLowImp}{-2.83} 
\newcommand{\mLowImpWid}{0.82}
\newcommand{\mHighImp}{-4.85} 
\newcommand{\mHighImpWid}{1.57}
\newcommand{\mbreakImp}{117} 
\newcommand{\mbreakImpWid}{36}
\newcommand{\nDownImp}{228} 
\newcommand{\mDownImp}{-3.47} 
\newcommand{\mDownImpWid}{1.42}
\newcommand{\nUpImp}{94} 
\newcommand{\mUpImp}{-4.22} 
\newcommand{\mUpImpWid}{1.39}

\title{Statistical results for solar energetic electron spectra observed over 12 years with STEREO/SEPT}

\correspondingauthor{Nina Dresing}
\email{dresing@physik.uni-kiel.de}

\author{Nina Dresing}
\affiliation{Institut fuer Experimentelle und Angewandte Physik \\
University of Kiel, Germany}

\author{Frederic Effenberger}
\affiliation{Helmholtz Centre Potsdam, GFZ German Research Centre for Geosciences, Telegrafenberg, Potsdam, 14473, Germany}
\affiliation{Bay Area Environmental Reserach Institute, NASA Research Park, Moffett Field, CA, 94043, USA}
\nocollaboration

\author{Ra\'ul G\'omez-Herrero}
\affiliation{Space Research Group, Universidad de Alcal\'a, Spain}
\nocollaboration

\author{Bernd Heber}
\affiliation{Institut fuer Experimentelle und Angewandte Physik \\
University of Kiel, Germany}
\nocollaboration

\author{Andreas Klassen}
\affiliation{Institut fuer Experimentelle und Angewandte Physik \\
University of Kiel, Germany}
\nocollaboration

\author{Alexander Kollhoff}
\affiliation{Institut fuer Experimentelle und Angewandte Physik \\
University of Kiel, Germany}
\nocollaboration

\author{Ian Richardson}
\affiliation{GPHI and Department of Astronomy,  University of Maryland, College Park, MD 20742,  USA and Heliospheric Physics Division, NASA Goddard Space Flight Center, Greenbelt, Maryland 20771, USA}

\author{Solveig Theesen}
\affiliation{Institut fuer Experimentelle und Angewandte Physik \\
University of Kiel, Germany}
\nocollaboration



\begin{abstract}
We present a statistical analysis of \ac{nr} solar energetic electron event spectra near 1 au. We use measurements of the STEREO \ac{sept} in the energy range of 45-425\,keV and utilize the \ac{sept} electron event list containing all electron events observed by STEREO~A and STEREO~B from 2007 through 2018. We select $\nall$ events with significant signal to noise ratios for our analysis and fit the spectra with single or broken power law functions of energy. We find $\ndpl$ ($\nspl$) events showing broken (single) power laws in the energy range of \ac{sept}. 
The events with broken power laws show a mean break energy of about 120\,keV.
We analyze the dependence of the spectral index on the rise times and peak intensities of the events as well as on the presence of relativistic electrons.
The results show a relation between the power law spectral index and the rise times of the events with softer spectra belonging to rather impulsive events.
Long rise-time events are associated with hard spectra as well as with the presence of higher energy ($>$0.7\,MeV) electrons.
This group of events cannot be explained by a pure flare scenario but suggests an additional acceleration mechanism, involving a prolonged acceleration and/or injection of the particles.
A dependence of the spectral index on the longitudinal separation from the parent solar source region was not found.
A statistical analysis of the spectral indices during impulsively rising events (rise times $<$ 20 minutes) is also shown.
\end{abstract}
\acresetall


\section{Introduction} \label{sec:intro}
While magnetic reconnection in solar flares is generally regarded as the mechanism that accelerates solar energetic electrons up to relativistic energies \citep[e.g.][]{Mann2015}, the role of other acceleration mechanisms such as \ac{cme}-driven shocks is still under debate. 
On the one hand, the existence of type II radio bursts demonstrates that \ac{cme}-driven shocks do accelerate electrons of a few keV \citep{Holman1983}. On the other hand, interplanetary shocks at 1\,au appear to be largely ineffective in accelerating \ac{nr} electrons, i.e. with energies of several tens to several hundreds of keVs \citep{Tsurutani1985, Dresing2016b, Yang2019}. 
Other cases of shock-accelerated electrons, e.g. at the Earth's bow shock \citep{Anderson1979}, the heliospheric termination shock \citep{Decker2005}, or at \acp{cir} \citep{Simnett2005} usually show only electrons up to around 100\,keV. 
Furthermore, model results of electron-shock acceleration \cite[e.g.][]{Guo2015, Trotta2019} show that energies of above 100\,keV can only be produced under certain conditions and by high-Mach-number shocks, which is supported by observations of MeV electrons apparently accelerated at Saturn's (high-Mach-number) bow shock \citep{Masters2013} where a main ingredient seems to be the presence of large amplitude magnetic fluctuations.
Favourable conditions for efficient electron acceleration at \ac{cme}-driven shocks might form at certain distances and positions along the shock front.
However, the presence of widespread electron events \citep{Dresing2014}, especially those which show prompt electron onsets and significant anisotropies over wide longitudinal ranges \citep[e.g.][]{Gomez-Herrero2015, Lario2016, Dresing2018} is not only hard to explain by a sole flare acceleration scenario but also (based on the above facts) by acceleration at a coronal or \ac{cme}-driven shock. 
During such widespread events, it is also not clear if the accelerator itself is extended or whether the interplanetary injection process, i.e. the particle release into \ac{ip} space, is responsible for the wide particle spread in these events. 
Therefore, mechanisms like the spreading of magnetic field lines close to the Sun \citep{Klein2008}, enhanced perpendicular transport, e.g. through field line meandering \citep{Laitinen2016} or particle scattering at magnetic irregularities \citep{Droege2003} may play an important role as well.
The presence of large-scale particle traps \citep{Dresing2018}, possibly caused by interacting \acp{cme}, may be a further mechanism not only able to explain wide particle injections but also efficient electron acceleration. \cite{Dresing2018} suggested that the presence of a large-scale particle trap may have provided the conditions for efficient electron acceleration during the 26 Dec 2013 widespread SEP event, which was also characterized by an extraordinary long-lasting electron injection, supposedly caused by continuous leakage from the trap.
\\
In a statistical study of 31 impulsive electron events observed with Wind/3DP \citep{Lin1995}, \cite{Krucker2007} compare the spectral index $\delta$ of the in-situ observed electrons with that of the \ac{hxr} photon spectrum of the corresponding flare. For a group of prompt events, they find a clear correlation of about unity suggesting that both the X-Ray emitting electrons and the in-situ electrons belong to the same source population. 
However, a second group of delayed events shows harder in-situ electron spectra and a much weaker correlation with the X-Ray spectrum, suggesting that the source spectrum may have been altered after the flare acceleration. This could be caused for example by on-going acceleration in post-flare loops \citep{Klein2005} or by re-acceleration in the CME-shock environment \citep{Petrosian2016}. 
Indeed, the common delays between the onset of interplanetary electron events and the solar injection time, inferred from the associated flare observation, as well as ramp-like, i.e., gradual rising phases during well-connected electron events, were attributed to CME-driven shocks \citep{Haggerty2002, Haggerty2009}.
\cite{Oka2018} summarize in their review on \ac{nr} electron spectra that reconnection processes, like solar flares or the Earth's magnetotail reconnection, will rather produce softer spectra  with spectral indices $\delta\sim$ 3-5 while shocks are known to produce harder spectra of $\delta\sim$ 2-4.
\\
A \ac{sep} spectrum observed in-situ in the \ac{ip} medium may also have been altered by various transport effects.
The generation of Langmuir waves by electrons of a few to tens of keV, for example, may cause energy loss and the formation of a spectral break. Model results by \cite{Kontar2009} yield a break energy of 35\,keV due to the spectral hardening of the lower energy component, whereas above the break, the spectrum resembles the  unchanged injection spectrum.
\\
Another transport effect may arise due to the energy-dependent diffusion coefficient describing the scattering of energetic particles at magnetic irregularities. Energy dependent transport modeling applied to different solar energetic particle events \citep{Droege2000, Agueda2014} shows that the mean free path of \ac{nr} electrons decreases with increasing energy. This means that the higher energy electrons experience stronger scattering than the lower energy ones with a constant mean free path for electrons above $\sim1$GV ($\sim1$600\,keV). \\
While the above transport effects may lead to a spectral break in the \ac{nr} electron spectrum, as is frequently observed during impulsive electron events \citep[e.g.][]{Lin1982, Reames1985, Krucker2009}, such a break could also be caused by the acceleration process. 
In particular, a typical feature of shock acceleration is a break or a roll-over corresponding to the acceleration limit caused by the finite extent of the shock front and/or the acceleration time \citep{Ellison1985}. 
However, the \ac{hxr} spectrum of a solar flare can also show a broken power law shape with a break energy of about 100\,keV as reported by \cite{Dulk1992}.
\\
In this work, we analyze the spectra of \ac{nr} electron events observed by the two STEREO spacecraft \citep{Kaiser2007} from 2007 until end of 2018. We do not separate or discard events depending on their rise times, intensities, delays or correlations to solar flares, \acp{cme}, type~III or type~II radio bursts.
To determine the spectra we use data of the \ac{sept} \citep{Muller-Mellin2007} with additional contributions from the \ac{het} \citep{Rosenvinge2008}.

\section{Event selection and data analysis}\label{sec:data_analysis}
Our study is based on the electron event list\footnote{\url{http://www2.physik.uni-kiel.de/stereo/downloads/sept_electron_events.pdf}} of \ac{sept} on board the two STEREO spacecraft. \ac{sept} measures electrons from 45-425\,keV in 15 energy channels and in four viewing directions.
The list is based on observations of 55-85\,keV electrons usually from the Sun telescope which looks along the nominal Parker spiral toward the Sun. However, if another of the four telescopes observes an earlier onset time, we use observations from this one. \\
For each event, the list provides the onset time, defined by the time when the intensity shows a 3-$\sigma$ increase above pre-event background. If the statistics allow, the original 1-min data is used, otherwise longer averages are applied. These averaging windows are taken as an estimate for the onset uncertainty and are also listed. Furthermore, the list provides the peak times and peak intensities, which are determined with at least 10-min averages or with the same averaging as used for the onset times, if longer averaging is applied. The peak time and intensity may be missing if another event, following shortly after, masks the event maximum.  A comment is provided and the events are numbered. 
\\
The list starts at the beginning of the STEREO science mission in January 2007 and events up to the end of 2018 are taken into account for the present study. 
Due to the loss of contact with STEREO~B, there are no STEREO~B events after Sep 2014.
The total number of events is 925, 557 at STEREO~A and 368 at STEREO~B.
For the present analysis all events of the list are taken into account regardless of the impulsiveness of the time profiles (defined by the rise time of the events), possible delays with respect to their solar counterpart, or the presence of radio type II and type  III bursts, or CMEs.\\
For each of the listed events, we determine the maximum intensity spectrum, taking the peak intensity of each energy channel based on 10 minute averaged data. We subtract the pre-event background for each energy channel, which includes background caused by preceding events, and use the telescope which observes the highest peak intensity. If no difference between the four viewing directions is observed, the Sun telescope is used.
Energy channels are excluded from the spectrum if their peak intensity does not increase significantly (3$\sigma$) above its pre-event background. We also exclude higher energy bins if the two neighboring lower bins were already excluded because of the significance level. Furthermore, we exclude energy bins if they show higher intensities than their lower energy neighbor, which would correspond to a power law with a positive slope. This is not expected for peak intensity spectra of solar energetic electron events and might be caused by another electron population mixing with the analyzed event or by instrumental effects.
Finally, if the real peak of an event is masked by the onset of another, more intense, event, we use the intensities detected directly before the new increase.
If the final spectrum contains less than four points (which implies a maximum energy of 80\,keV) or the fit results is unreliable based on very high reduced chi-square values, the event is excluded from our analysis.
After applying the above restrictions $\nall$ events are left which are used in a statistical analysis.\\
We use the scipy.odr package of python to fit the data including uncertainties in intensity and energy, where the width of the energy bins serves as uncertainty.
To characterize each spectrum, we first assume a broken power law shape
\begin{equation}
\frac{dJ}{dE}\propto
\begin{cases}
   E^{\delta_1} & E < E_b\\
   E^{\delta_2} & E > E_b
\end{cases}
\end{equation}
where $\delta_1$ ($\delta_2$) denotes the spectral index below (above) the break energy $E_b$.
The fitting procedure is as follows: 
For a set of assumed break point positions we fit two single power law functions to each part of the dataset separated by the assumed break point. This set of assumed break points is determined by the energy binning of \ac{sept} and the restriction that at least three energy bins have to lie on each side of the assumed break point. After performing fits for the whole set of potential break points, the fit yielding the smallest difference between resulting and assumed break point is chosen to be the best. The break energy is determined by the intersection of the two single power law fits. Afterwards we also fit a single power law to the whole spectrum and chose the best of the two (single or broken power law) based on their reduced chi squares.
The \ac{sept} energy range (45-425\,keV) together with our fitting procedure limits the positions of break points, which can be determined, to the energy range between $\sim$70 and $\sim$300\,keV. 
\\
Due to its measurement principle (the magnet-foil technique) SEPT's electron measurements can be subject to ion contamination \citep{Wraase2018}:
The electron telescopes are covered by a thin ($4.95 \mu$m) Parylene foil which leaves the electron spectrum essentially unchanged  but  stops  ions  of  energies  up  to $\sim$400\,keV.  Ions  with  energies  above  this  limit  can cross the foil and stop inside the top silicon detector, being registered as electrons by the sensor. This effect, mainly caused by protons of a few hundred keV but below 1\,MeV, is well known and can be identified as described in the instrument caveats\footnote{\url{http://www2.physik.uni-kiel.de/stereo/data/sept/level2/SEPT_L2_description.pdf}}. 
A further potential contamination effect caused by electrons and protons in the MeV-range, which are able to enter the telescope from the side.
This is not identified so straightforwardly in the data but can be corrected by using simulated response functions of the telescope \citep[see][]{Wraase2018, Kollhoff2019}.
We apply a contamination correction to the level2-electron data consisting of three parts: i) protons in the hundreds of keV range, determined by the SEPT proton measurements, ii) relativistic electrons, determined using the measured spectra of the \ac{het} electrons from 0.7-4\,MeV and iii) high-energy protons, determined by HET proton measurements from 13 to 100\,MeV.
The input spectra for this correction were determined by using the measured spectra of the respective particles in the same time window used to determine the peak intensity spectrum at SEPT. 
Because of the upper energy limit of the HET measurements, the input spectra for the correction were extrapolated to 10\,MeV for electrons and 1\,GeV for protons assuming a single power law shape.
Furthermore, the electron input spectrum was multiplied by an intercalibration factor of 14 as found by \cite{Richardson2014} comparing the HET with SOHO/EPHIN electron intensities. 
Finally, the contamination correction is also used to exclude \ac{sept} energy bins from the spectrum if the contamination is larger than 50\%. 
For energy bins suffering a smaller but non-zero contamination, we propagate the error of the contaminating intensities to the corrected intensities.
%
Our list of electron events was also synchronized with the $>$25\,MeV proton event list by \cite{Richardson2014} and a non-published extended version of that list until April 2017.

\section{Results}\label{sec:results}
\subsection{Statistical results}
\begin{figure}
\plotone{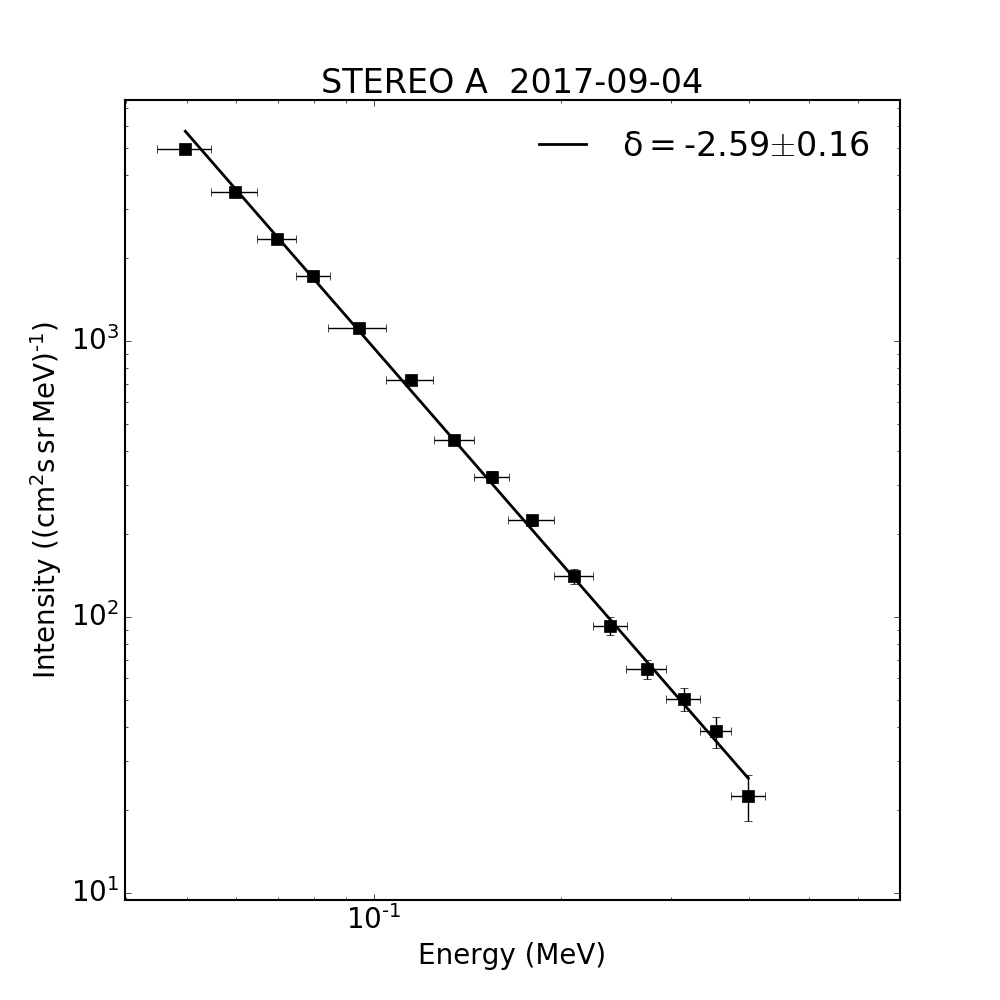}
\caption{Example of an event with a single power law peak spectrum.}
\end{figure}\label{fig:example_single_pl}
\begin{figure}
\plotone{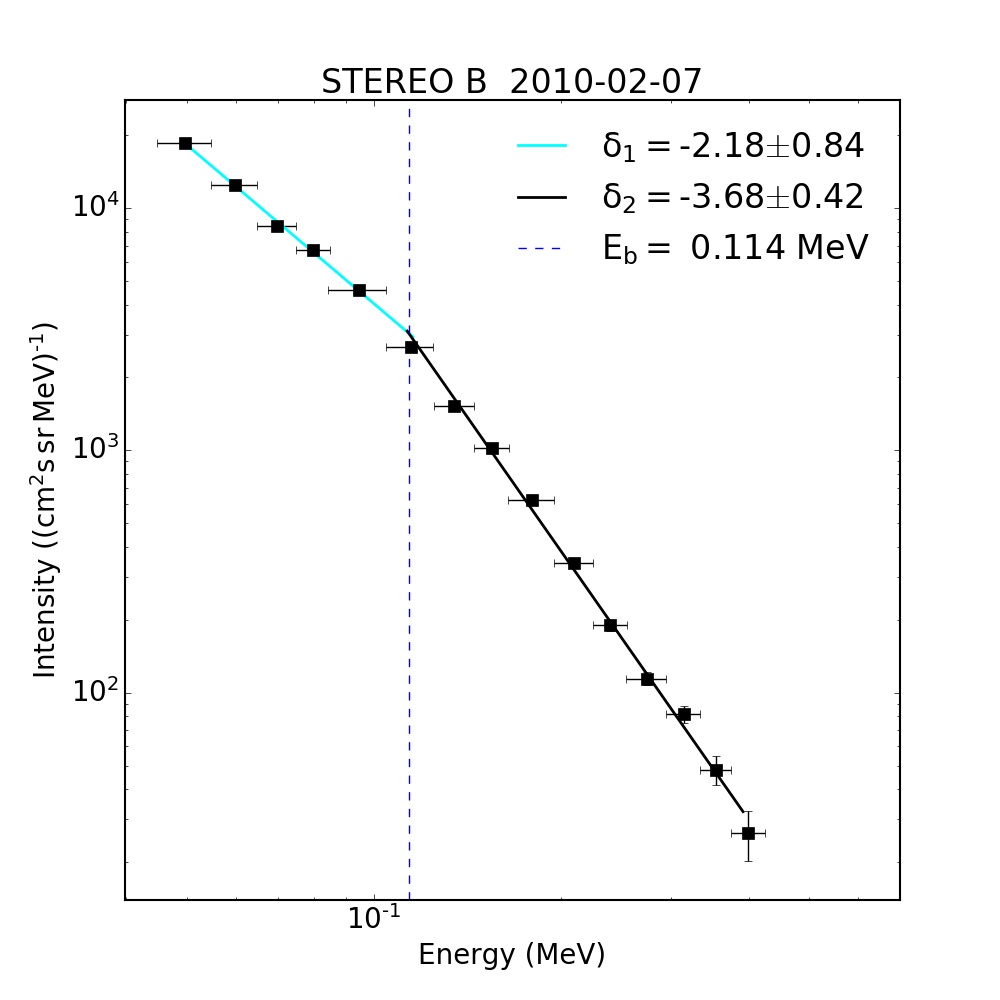}
\caption{Example of an event with a broken power law peak spectrum.}
\end{figure}\label{fig:example_broken}
\begin{figure*}
\plotone{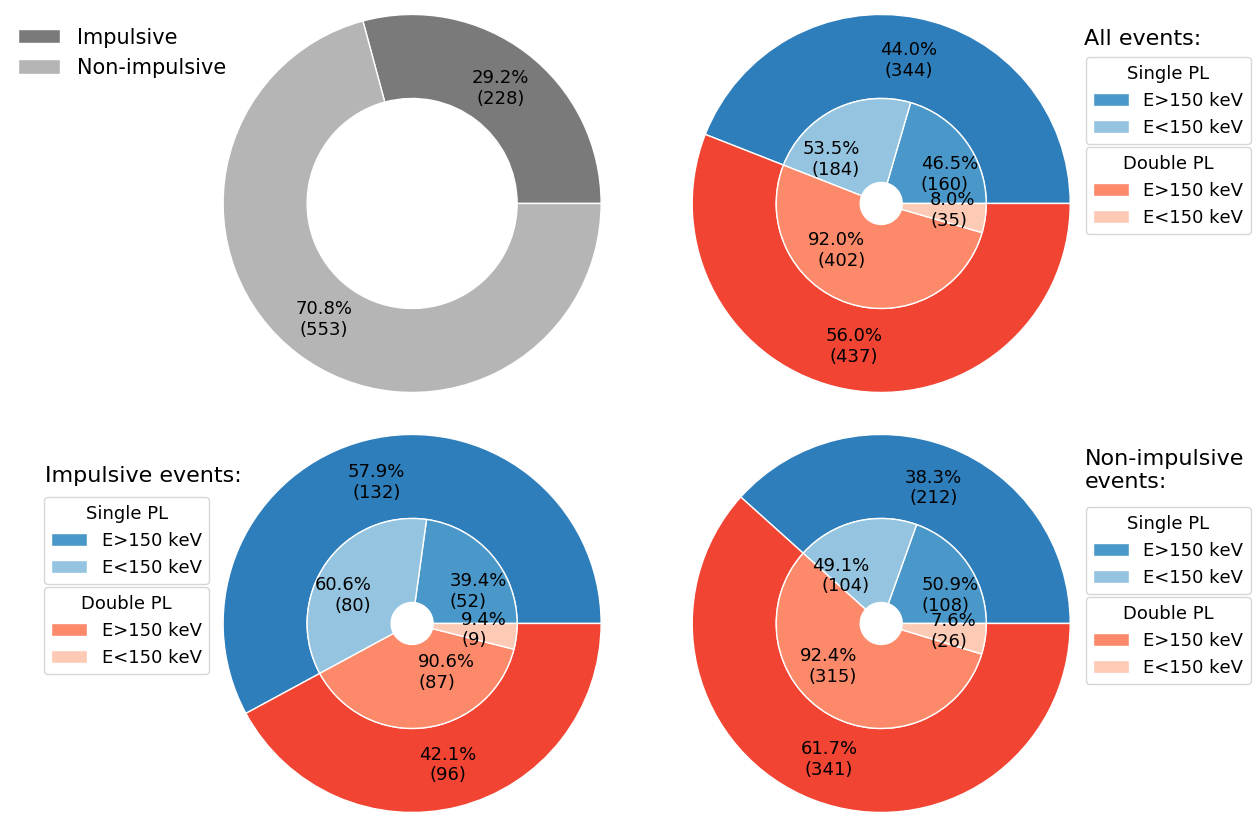}
\caption{Statistical results for electron event properties. Top left: Relative numbers of events with impulsive ($<20$\,min) and non-impulsive rise times. Top right: The outer ring shows the relative numbers of events with single power shape (blue) and broken power law shapes (red). The inner ring further separates the events into sub-categories of events detected up to a maximum energy above or below 150\,keV (lighter colors belong to the corresponding darker color of the outer ring).
The bottom panels are in the same format as the top right panel, but show results for the sub samples of impulsive (left) and non-impulsive (right) events.}
\end{figure*}\label{fig:donut}
\begin{figure*}
\plottwo{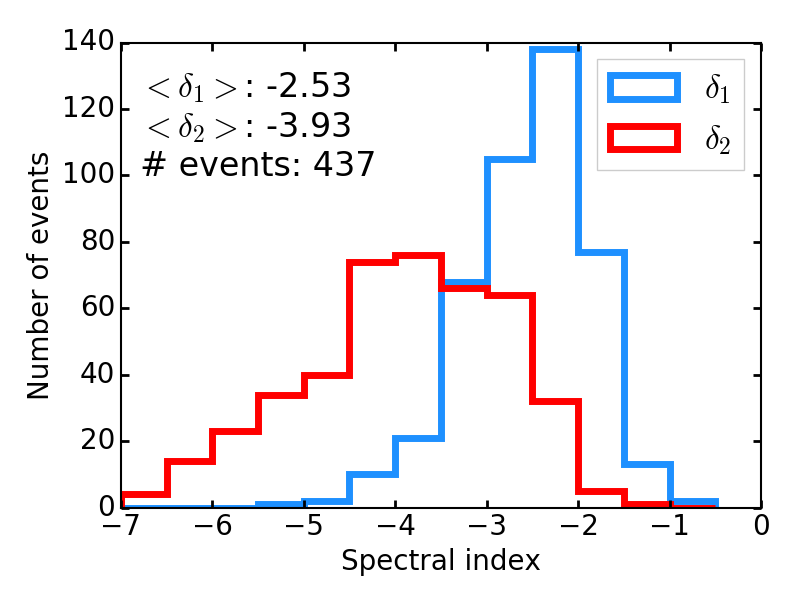}
{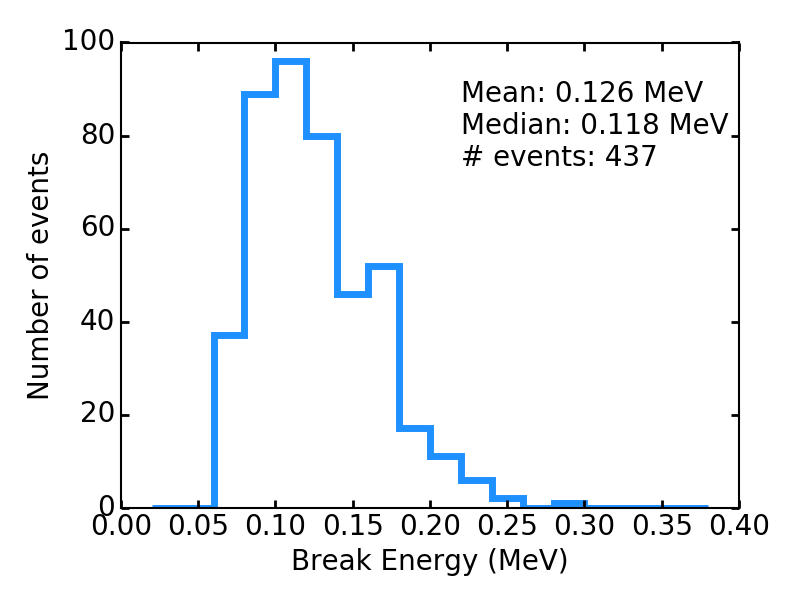}
\caption{Spectral indices for broken power law events (left) and corresponding break energies (right).}
\end{figure*}\label{fig:hist_gamma1_2_break}
\begin{figure*}
\plottwo{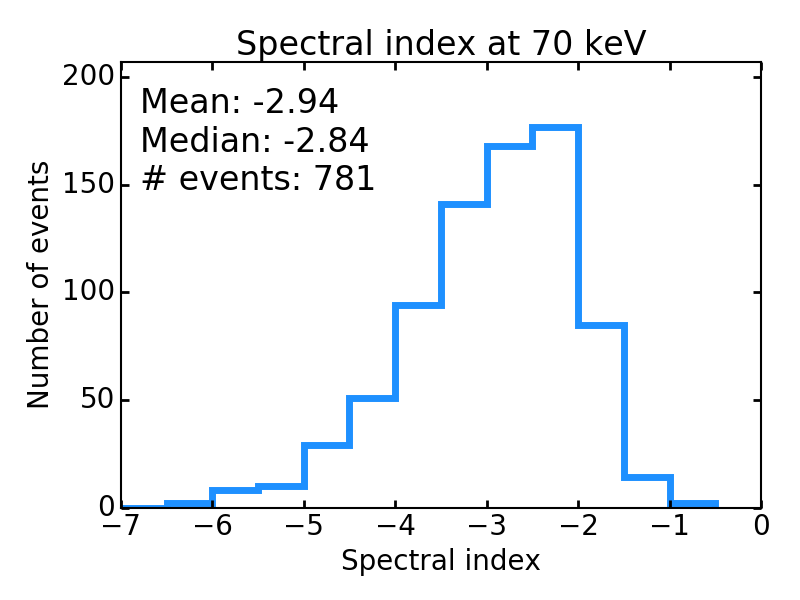}
{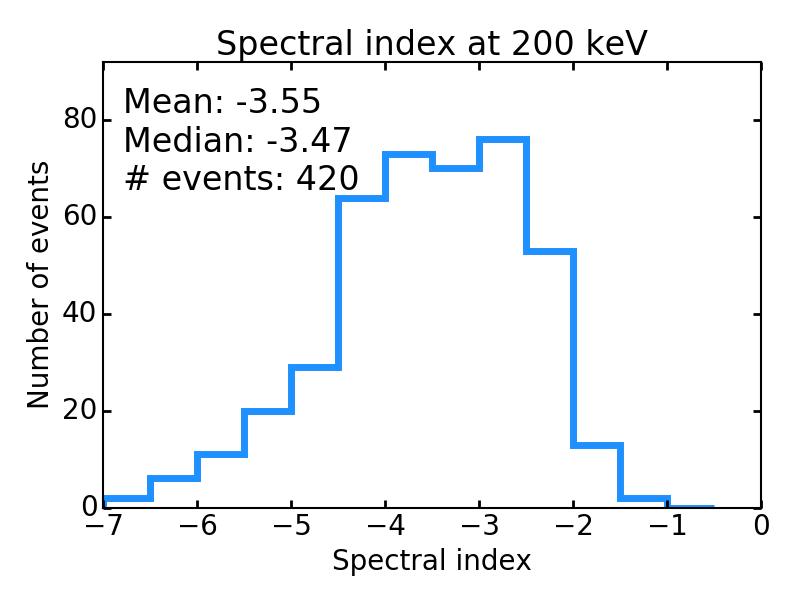}
\caption{Spectral indices at 70\,keV (left) and 200\,keV (right) for the whole sample of events.}
\end{figure*}\label{fig:hist_gamma70_200}
The analysis of the whole sample of electron events reveals different spectral shapes. 
While some events ($\nspl$) show a single power law shape like the one shown in Fig. \ref{fig:example_single_pl}, others ($\ndpl$) are described by a broken power law (see Fig. \ref{fig:example_broken}) which is the typical shape previously reported in the analyzed energy range especially for impulsive solar energetic electron events  \citep[e.g.][]{Krucker2009}. 
As a sub-sample of all of the \ac{sept} electron events, we select those events with impulsive rise times which we define as showing delays between onset and peak time smaller than 20 minutes in the energy range of 55-85\,keV. This criterion is based on the distribution of rise times discussed below in relation to Fig. \ref{fig:hist_ttm}
Note, that we use the term "impulsive" here only with regards to the time profiles and it does not imply that the event has the characteristics (such as ion compositional signatures) of an "impulsive" solar particle event \citep[e.g.][]{Reames1999}.
Figure \ref{fig:donut} (top-left) shows that 29\% of the events were impulsive, while the majority were non-impulsive with rise times $>20$ minutes.
The top-right diagram of Fig. \ref{fig:donut} shows, in the outer ring, that the event spectra were nearly evenly divided between single power laws (blue; 44\%) and broken power laws (red; 56\%).
The outer rings of the bottom left and right panels of Figure \ref{fig:donut} similarly summarize the fractions of single and broken power law spectra for impulsive and non-impulsive events.
The inner rings in Fig. \ref{fig:donut} represent the relative numbers of events with maximum energies above or below 150\,keV. 
The maximum energy is the mean energy of the highest energy bin used in the fit and is usually determined by the highest energy bin showing a 3-$\sigma$ increase above its pre-event background. 
However, if energy bins are excluded because of proton or high energy electron contamination, the maximum energy may be under-estimated.
The top-right and bottom diagrams show that out of the single power law events (blue), $\gtrsim$50\% have maximum energies $<150$\,keV implying that a break point at energies $\gtrsim100$\,keV would have been missed.
Regarding the single power-law events reaching higher energies, one possibility is that a break point was present but detection was still limited by the observations and the fitting procedure, which requires at least three points on each side of the break. In this case, the break could have been below 70\,keV or above 300\,keV.  Alternatively, the spectrum could in fact have been a single power law.
\\
Fig. \ref{fig:hist_gamma1_2_break} shows histograms of the spectral indices (left) and the break energy (right) for events showing a double power law shape. 
The mean spectral index below (above) the break is $\delta_1=\mLow\pm\mLowWid$ ($\delta_2=\mHigh\pm\mHighWid$) with the distribution of $\delta_2$  being wider. Note that the uncertainties provided above represent the widths of the distributions determined by the span between the first and third quartiles of the distributions. Because of the non-Gaussian shapes of several of the distributions we prefer not to use a 1-$\sigma$ uncertainty as provided by \cite{Krucker2009} for their mean spectral indices which yield smaller values. 
The mean spectral index for the events showing only a single power law lies between $\delta_1$ and $\delta_2$ with $\delta=\mspl\pm\msplWid$ (not shown). 
For comparison, \cite{Krucker2009} find values of $\delta_1=-1.9$, $\delta_2=-3.6$ for a sample of 62 impulsive electron events observed in solar cycle 23. While the values for $\delta_2$ agree reasonably well, \cite{Krucker2009} found a harder  $\delta_1$ and a smaller break energy.
Especially, when comparing our values of $\delta_1$ and $\delta_2$ of the impulsive sample (see Table \ref{tab:means}) we find significantly softer values than \cite{Krucker2009} which might be caused by the weaker solar cycle 24 but could also be due to the different instrumentation and their non-equal energy ranges used in the studies.
We find a mean break energy of $E_b=\mbreak\pm\mbreakWid$\,keV while \cite{Krucker2009} find a break energy of about 60\,keV. 
Note, however, that the \ac{sept} measurements and our fitting procedure only allow to determine a spectral break between $\lesssim 70$\,keV and $\gtrsim 300$\,keV which will presumably also impact the range of $\delta_1$. 
\\
To be able to analyze all events in our sample together, regardless of their spectral shape or the position of the eventual break point, we define two spectral indices $\delta70$ and $\delta200$ corresponding to the spectral index found in the energy range around 70\,keV and 200\,keV, respectively. In the case of single power law events $\delta70$ and $\delta200$ have equal values. For double power law events, defined by the two spectral indices $\delta_1$ and  $\delta_2$, we define $\delta70$ and $\delta200$ as follows:

\begin{eqnarray*}
\delta70 = \delta_1 \quad &:E_b& > 70 \mathrm{keV}\\
\delta70 = \delta_2 \quad &:E_b& < 70 \mathrm{keV}\\
\delta200 = \delta_1 \quad &:E_b& > 200 \mathrm{keV}\\
\delta200 = \delta_2 \quad &:E_b& < 200 \mathrm{keV}
\end{eqnarray*}
These two energies, 70\,keV and 200\,keV, lie outside most of the break energies for the broken power-law events (cf. Fig. \ref{fig:hist_gamma1_2_break} right) so that for most of those events $\delta70$ ($\delta200$) represents $\delta_1$ ($\delta_2$).
The corresponding distributions of $\delta70$ and $\delta200$ are shown in Fig. \ref{fig:hist_gamma70_200}. 
The mean spectral indices are  $\delta70=\mDown\pm\mDownWid$ and $\delta200=\mUp\pm\mUpWid$. The number of events for $\delta70$ is larger because many events do not extend to energies of 200\,keV.
Table \ref{tab:means} summarizes the mean values of the spectral indices of $\delta70$ and $\delta200$, the single power law events ($\delta$), the broken power law events ($\delta_1, \delta_2$) as well as the corresponding break points $E_b$. 
The table also provides the widths of the distributions.
\begin{deluxetable}{l|cccccc}
\tablecaption{Means of spectral indices and break energies for all events and the impulsive sample.\label{tab:means}}
\tablehead{
\colhead{} & \multicolumn{3}{c}{All events} & \multicolumn{3}{c}{Imp. sample}\\
\colhead{} & \colhead{Mean} & \colhead{Width\tablenotemark{*}} & \colhead{$\#$} & \colhead{Mean} & \colhead{Width} & \colhead{$\#$}
 }
\startdata
Total $\#$ & & & $\nall$ & & & $\nallImp$ \\
$\delta$ (single PL) & $\mspl$ & $\msplWid$ & $\nspl$  & $\msplImp$ & $\msplImpWid$ & $\nsplImp$ \\
$\delta_1$ & $\mLow$ & $\mLowWid$ &  $\ndpl$ & $\mLowImp$ & $\mLowImpWid$ &  $\ndplImp$\\
$\delta_2$ & $\mHigh$ & $\mHighWid$ &  $\ndpl$  & $\mHighImp$ & $\mHighImpWid$ & $\ndplImp$ \\
$E_b$ & $\mbreak$\,keV & $\mbreakWid$\,keV & $\ndpl$  & $\mbreakImp$\,keV & $\mbreakImpWid$\,keV & $\ndplImp$\\
$\delta70$ & $\mDown$ & $\mDownWid$ & $\nDown$ & $\mDownImp$ & $\mDownImpWid$ & $\nDownImp$ \\
$\delta200$ & $\mUp$ & $\mUpWid$ & $\nUp$ & $\mUpImp$ & $\mUpImpWid$ & $\nUpImp$ \\
\enddata
\tablenotetext{*}{The width is determined by the span between the first and third quartiles of the distributions.}
\end{deluxetable}
\begin{figure*}
\plottwo{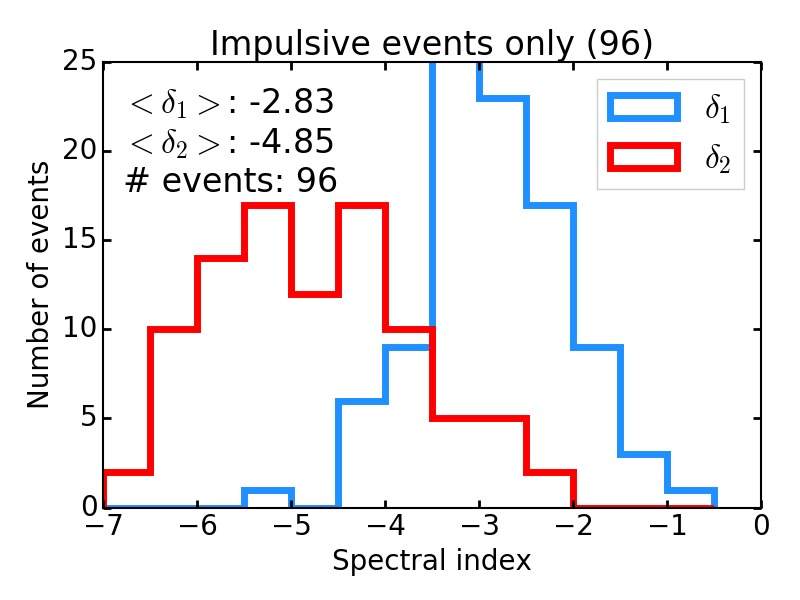}
{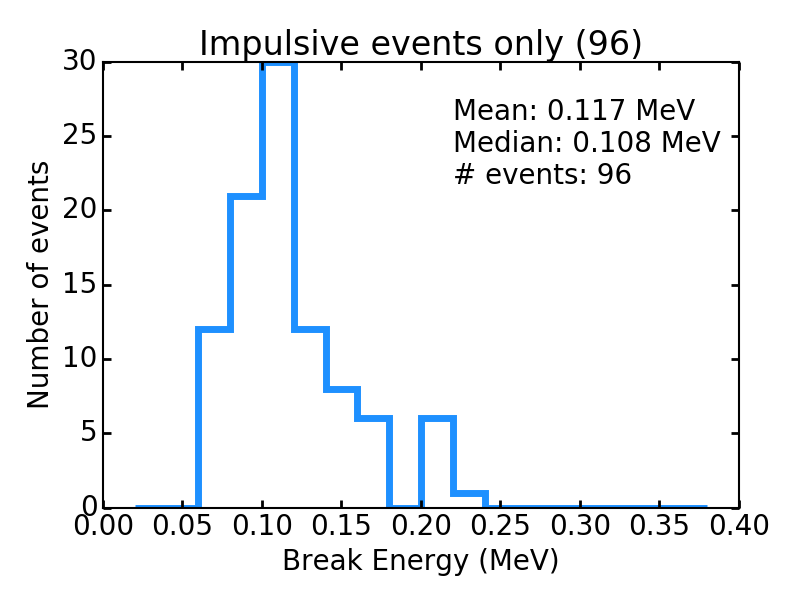}
\caption{Histogram of power law indices (left) and break points (right) for the impulsive sample.
}
\end{figure*}\label{fig:hist_gamma1_2_break_imp}

\begin{figure*}
\plottwo{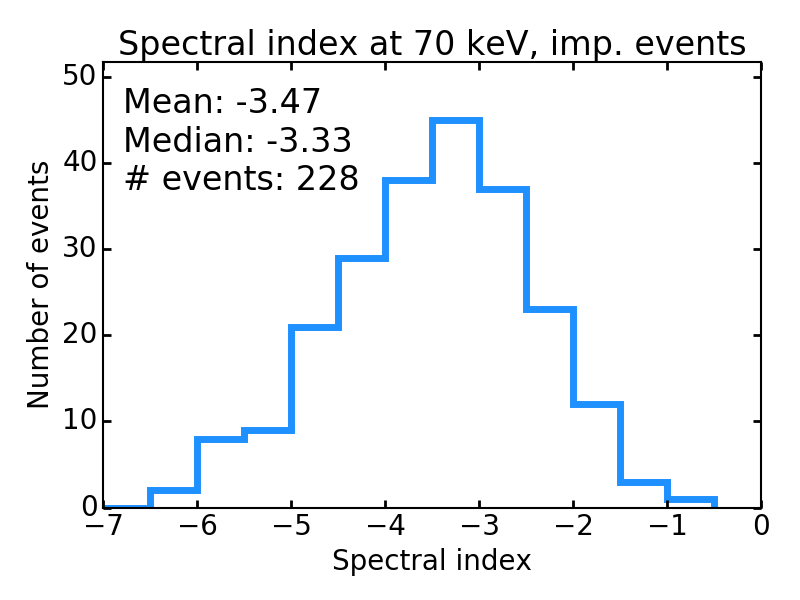}
{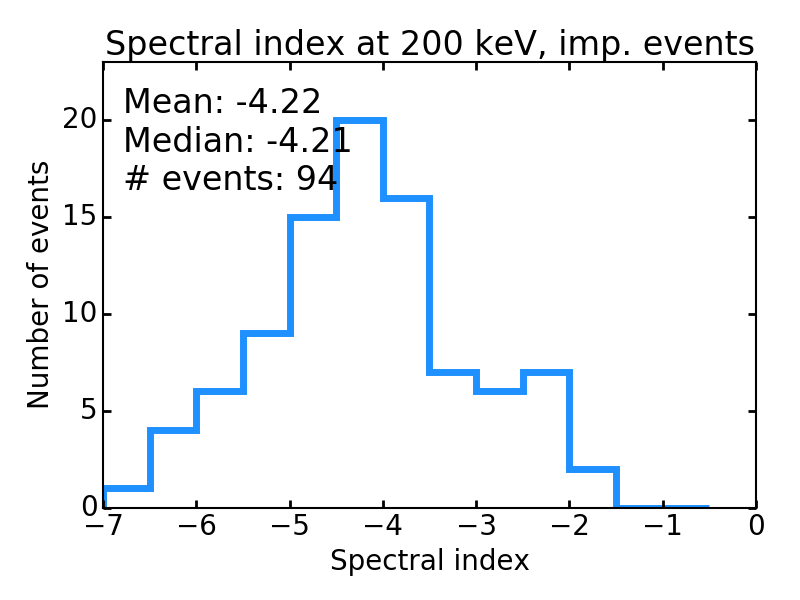}
\caption{Histogram of spectral indices at 70\,keV and 200\,keV for the impulsive sample.}
\end{figure*}\label{fig:hist_gamma70_200_imp}
Figures \ref{fig:hist_gamma1_2_break_imp} and \ref{fig:hist_gamma70_200_imp} show the same distributions of spectral indices and break energies as in Figures \ref{fig:hist_gamma1_2_break} and \ref{fig:hist_gamma70_200} but for the sub sample of impulsively rising events.
The mean spectral indices and break points of the impulsive sample are also listed in Table \ref{tab:means}. The break point is slightly smaller and all the spectral indices are on average softer when compared to the whole sample.
The largest difference applies to $\delta_2$ with a mean spectral difference of 0.85. The smallest average change of 0.33 is observed for $\delta_1$. The distributions of spectral indices are always wider for the impulsive sample.
%
%
%
\subsection{Correlations of spectral features and event properties}
\label{sec:correlation}
%
%
For the broken power law events Fig. \ref{fig:gamma_1_vs_2} plots the spectral indices above and below the break against each other with all events on the left hand side and only impulsive events on the right hand side.
Note, that the few outlier events (open symbols), which correspond to an unexpected spectral hardening above the break, have been excluded from the fit and the correlation (provided in the legend).
Most of these are events occur during periods of ion contamination and the spectral hardening could be due to an under-correction of the contamination effect.
The correlation between $\delta_1$ and $\delta_2$ is smaller for the impulsive sample, however, this could be due to the poorer statistics.
Nevertheless, we find a clear linear correlation between the spectral indices of 0.69 which in agreement to previous results \citep{Krucker2009} finding a correlation coefficient of 0.61.\\
Fig. \ref{fig:gamma_diff} displays the difference between the spectral indices below and above the break. 
The mean difference is larger for the impulsive sample (1.8) which compares better to the results by \cite{Krucker2009} who find a mean difference of 1.7 for their sample of 62 impulsive events.
\begin{figure*}
\plottwo{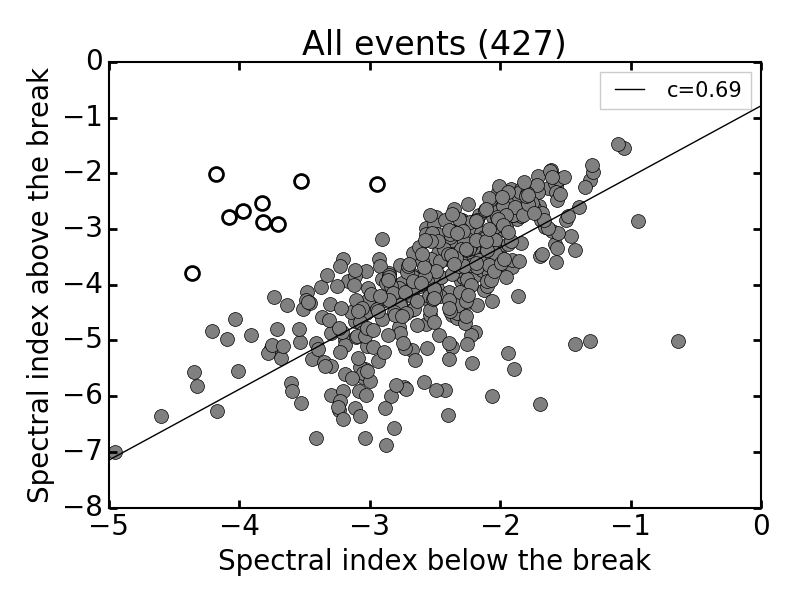}
{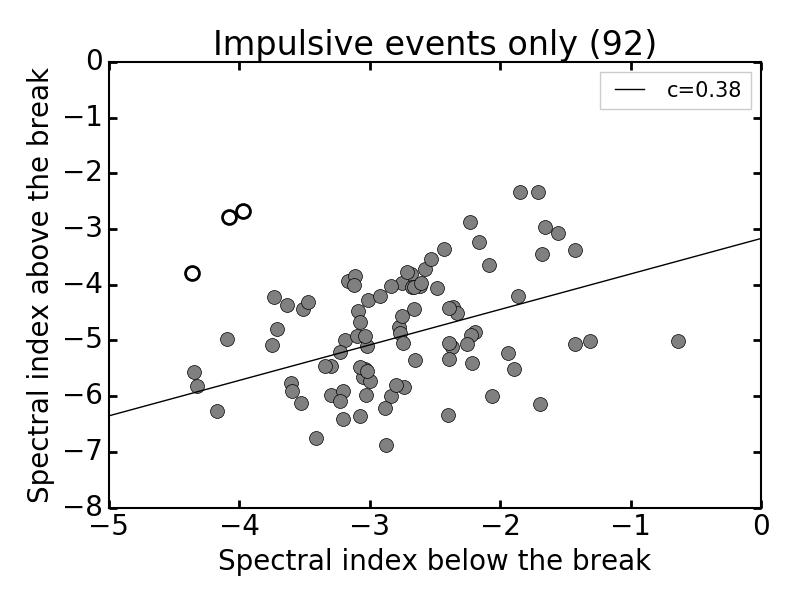}
\caption{Spectral index above the break as function of the spectral index below the break for all broken power law events. Left: All events in the sample, right: only events with impulsive rises. The black line represents a linear regression and the Pearson correlation coefficients are provided in the legend. Note, that events showing a spectral hardening above the break (open symbols) have been excluded from the correlation.}
\end{figure*}\label{fig:gamma_1_vs_2}
\begin{figure*}
\plottwo{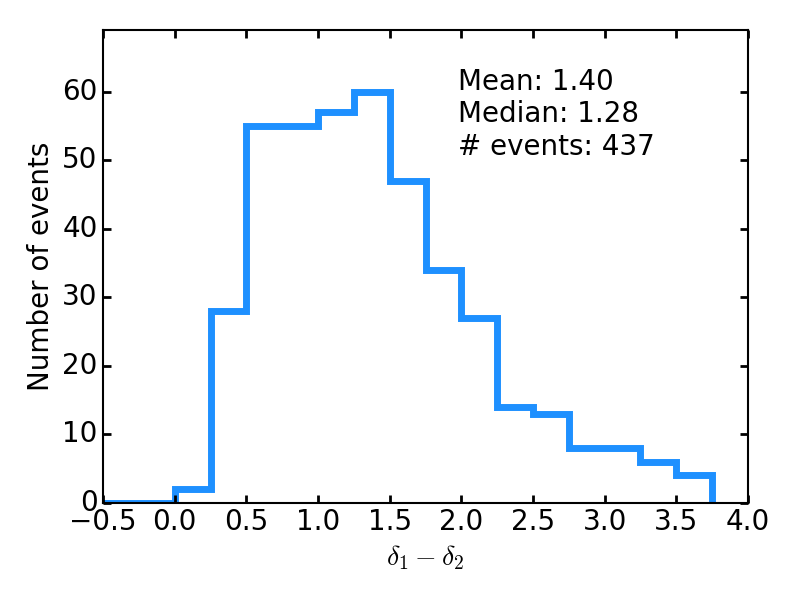}
{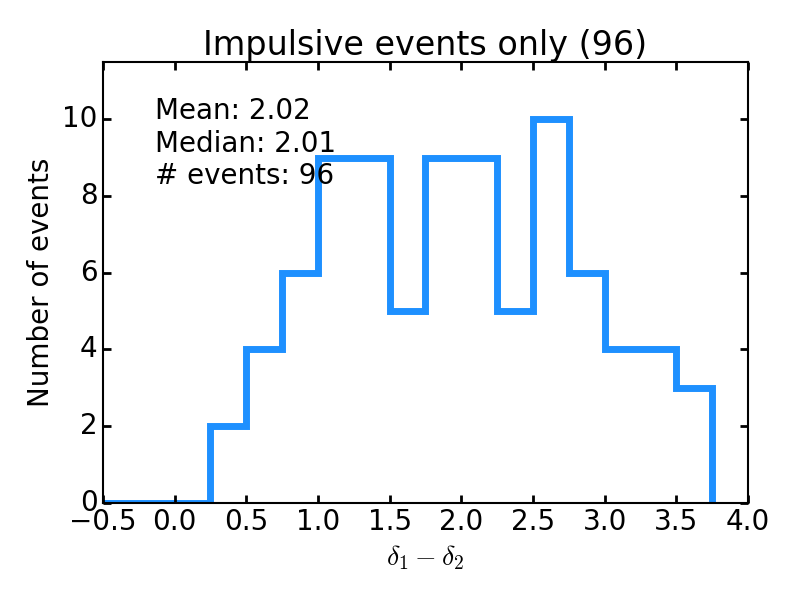}
\caption{Difference between spectral index below the break and above the break for all events (left) and impulsive events only (right).}
\end{figure*}\label{fig:gamma_diff}
\begin{figure}
\plotone{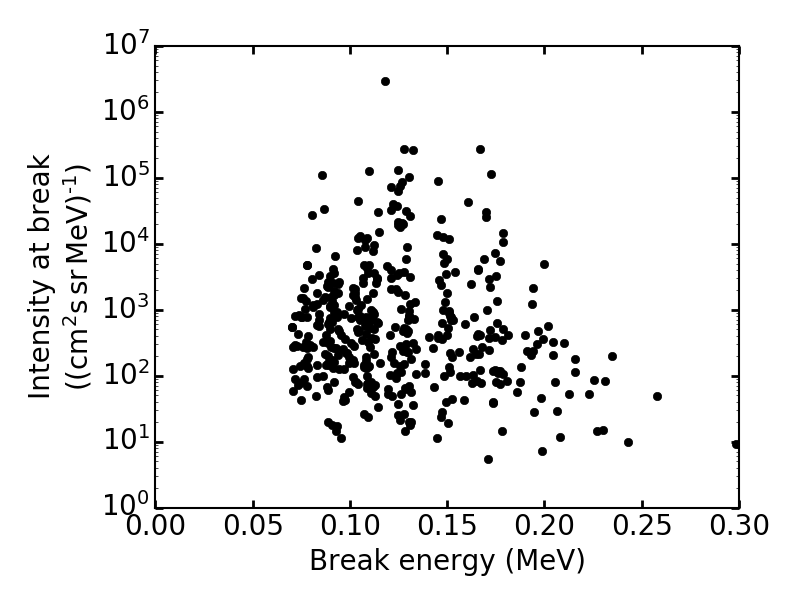}
\caption{Break energy as a function of the intensity at the break energy (pre-event background subtracted) for all broken-power law events.}
\end{figure}\label{fig:break_vs_int_at_break}
\\
Fig. \ref{fig:break_vs_int_at_break} plots the break energy of the broken power law events as a function of the intensity at the break energy and shows that there is no correlation. The values of break energies are rather equally distributed around the mean value. The grouping of points in columns reflects the limitation of the break-point determination caused by the energy binning of the instrument.
%
\begin{figure*}
\plotone{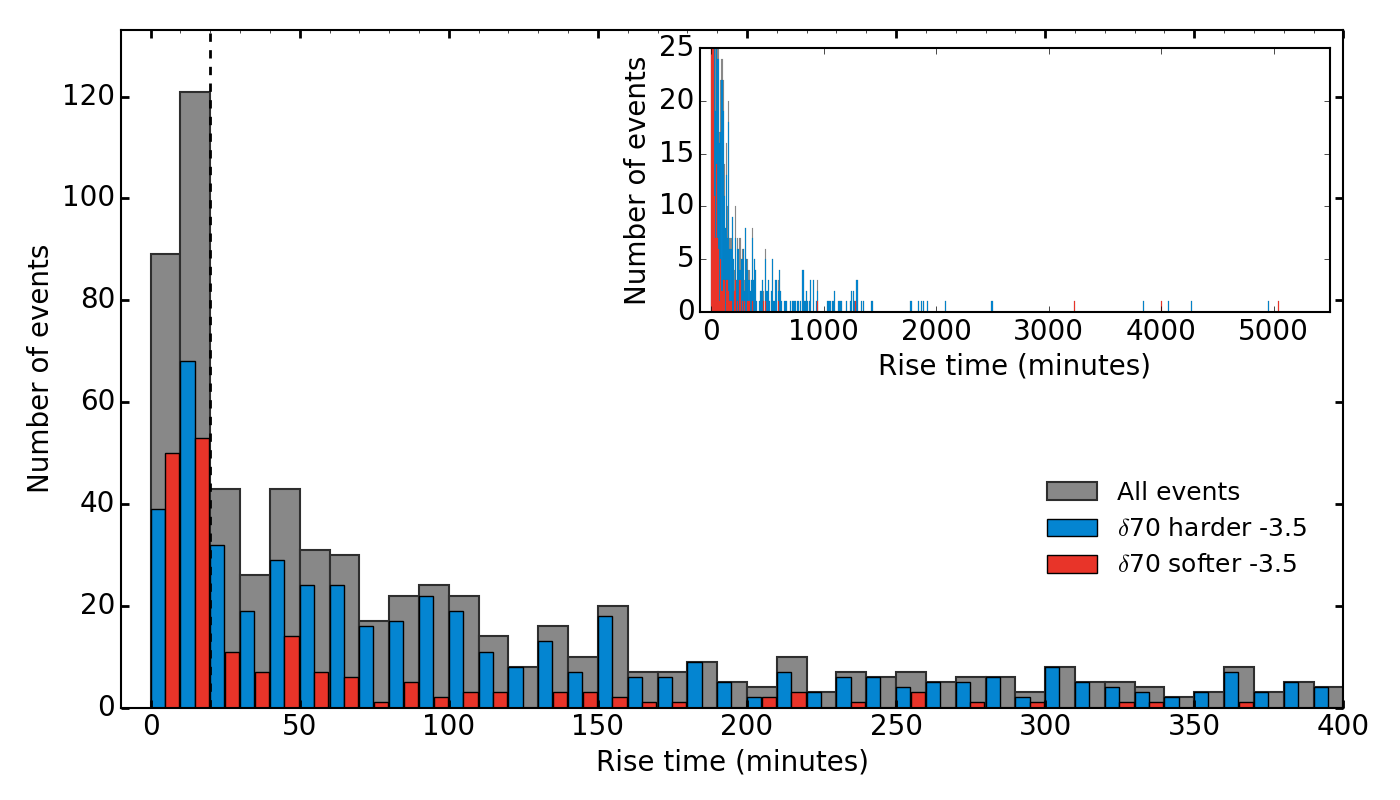}
\caption{Histogram of rise times of all events (gray) with a 10-minute binning zoomed in to 0-400 minutes.
The colored histograms divide the sample of all rise times based on the correspondingly measured spectral index $\delta70$. The number of events with harder spectra ($\delta70>-3.5$) is plotted in blue and the one with softer spectra ($\delta70<-3.5$) in red. The inset at the top right shows the histograms zoomed out on the x-axis but zoomed in on the y-axis where the histogram of soft spectra events (red) is plotted on top of the one with the hard spectra events (blue) and the histogram of all events (gray) in the back.}
\end{figure*}\label{fig:hist_ttm}
\begin{figure}
\includegraphics[width=0.5\textwidth, trim=0 12.2cm 0 0, clip] {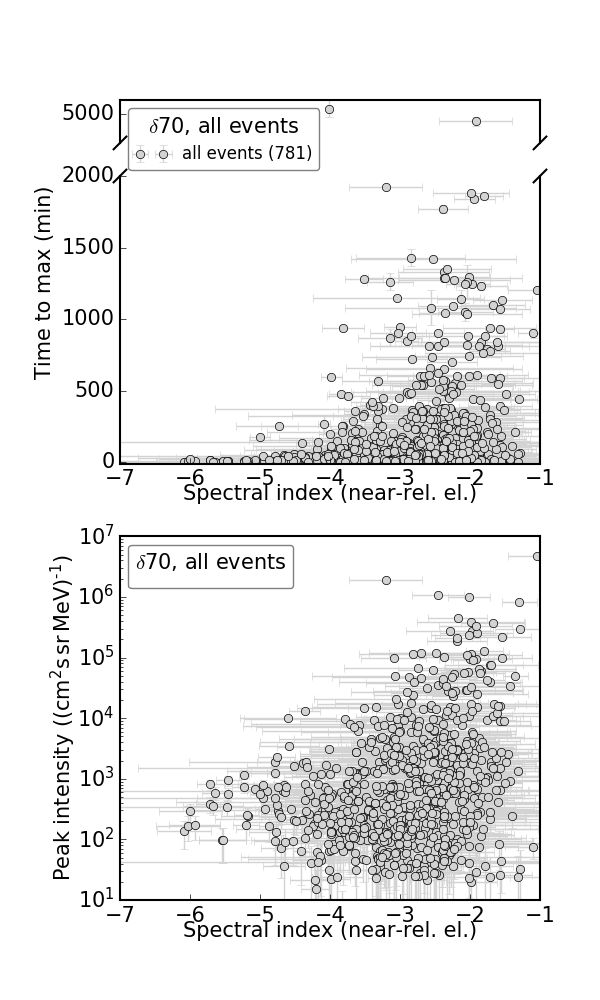}
\caption{Rise times (time from onset to maximum) of all 55-85\,keV electron events in the SEPT electron event list as a function of the spectral index at 70\,keV ($\delta70$).}
\end{figure}\label{fig:ttm_vs_gamma}
\\
Figure \ref{fig:hist_ttm} shows the histogram of rise times (times from onset to peak time) with 10 minute binning in gray.
While the main plots zooms in to rise times between 0-400 minutes, the inset shows the whole histogram, revealing that a few events show rise times of even some days. A few of these extremely long rising events are accompanied by a \ac{cme}, driving a shock, with comparable time scales of the electron rise and the \ac{cme} propagation suggesting that the \ac{cme}-driven shock could be the main source for those events.
However, most of the very long rise time events are showing only very small intensity increases and are not accompanied by an in-situ passage of a shock.
The colored bars show the number of events in the corresponding bin, which have a spectral index $\delta70$ softer (red) or harder (blue) than -3.5.
While softer events cluster more at small rise times ($<20$ min), however constituting only about half of all events, the harder spectra events dominate at longer rise times. 
Above the second bin ($>20$ min), the number of events per bin is significantly smaller and decreases gradually. We, therefore, choose the limit for impulsive and non-impulsive events (see Section \ref{sec:results}) to be at 20 minutes, marked by the dashed line.
\begin{figure*}
\plottwo{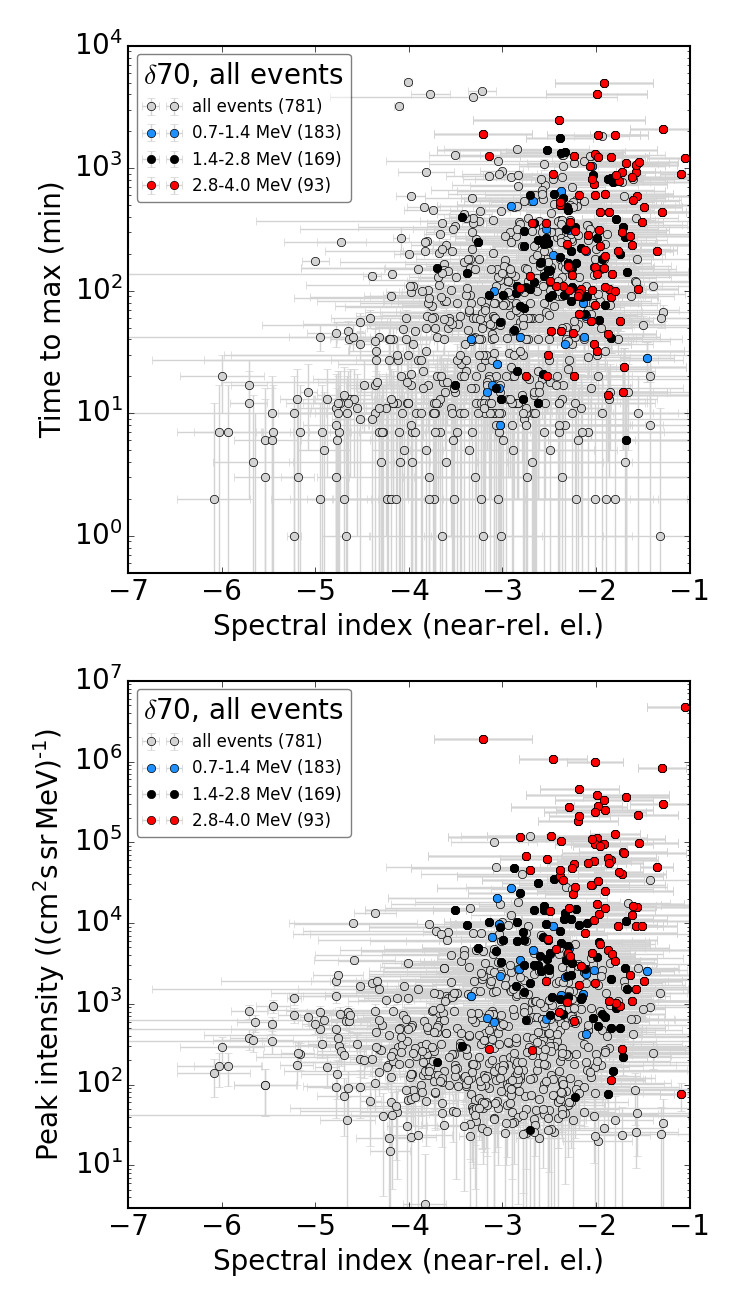}
{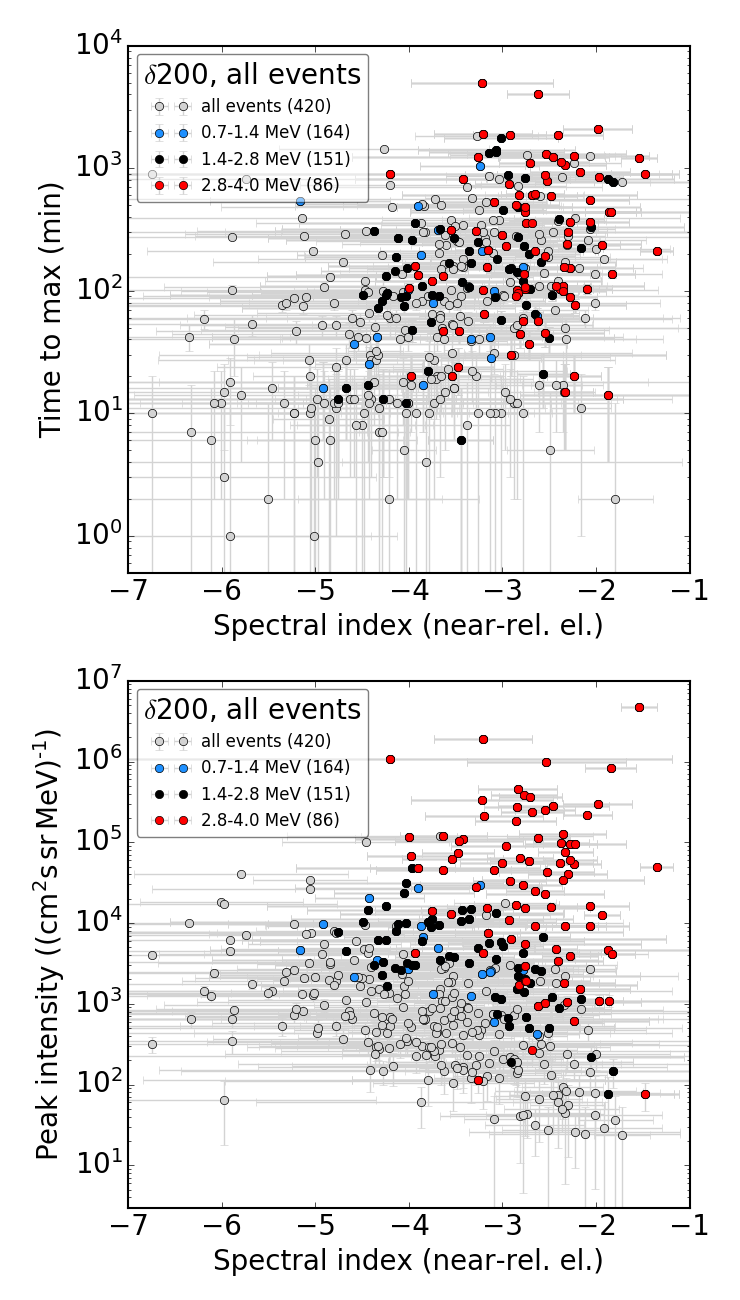}
\caption{Rise times (top) and peak intensities (bottom) at 55-85\,keV (pre-event background subtracted) as a function of the spectral index (gray points) at 70\,keV (left) and 200\,keV (right). The colored points mark those events where electrons in the MeV range (measured by STEREO/HET) were also present with red denoting that the highest HET channel (2.8-4\,MeV) was populated, black meaning that only energies up to the second channel (1.4-2.8\,MeV) were observed, and blue only in the first HET channel (0.7-1.4\,MeV), respectively.}
\end{figure*}\label{fig:gamma_vs_ttm_Imax}
\begin{figure*}[b!]
\plottwo{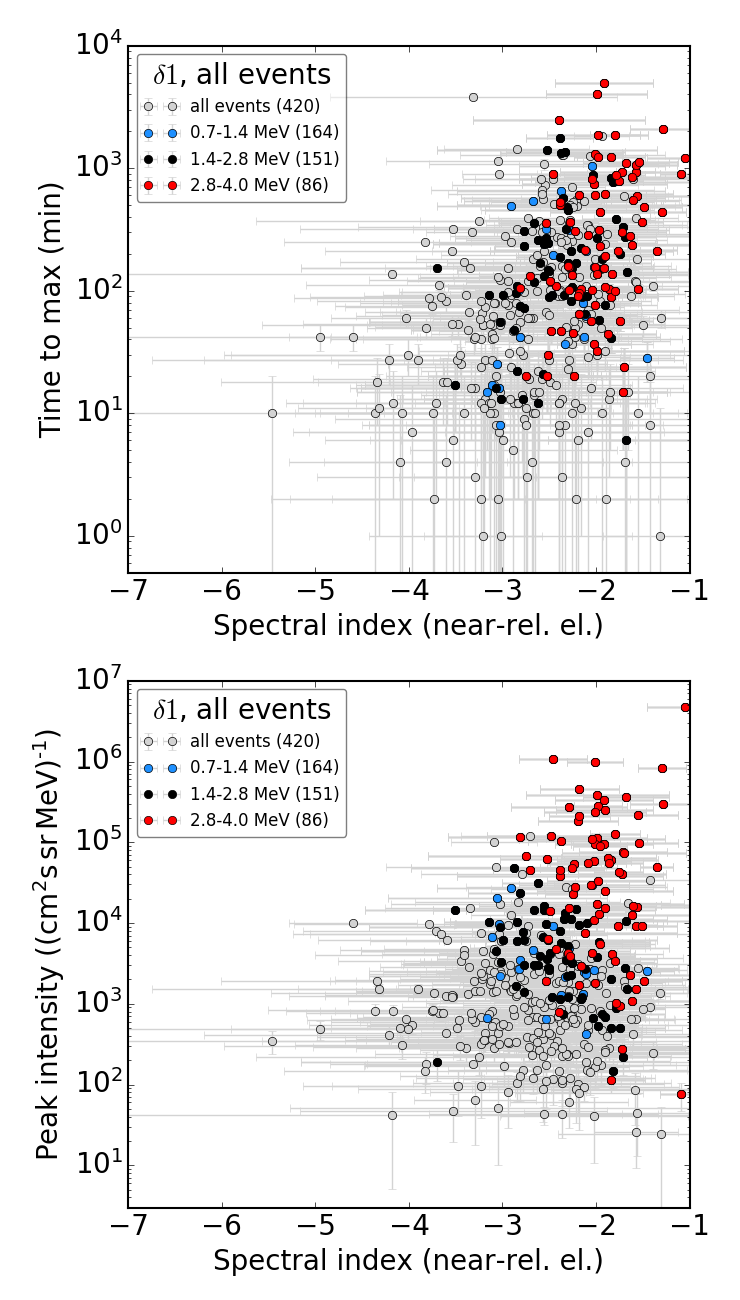}
{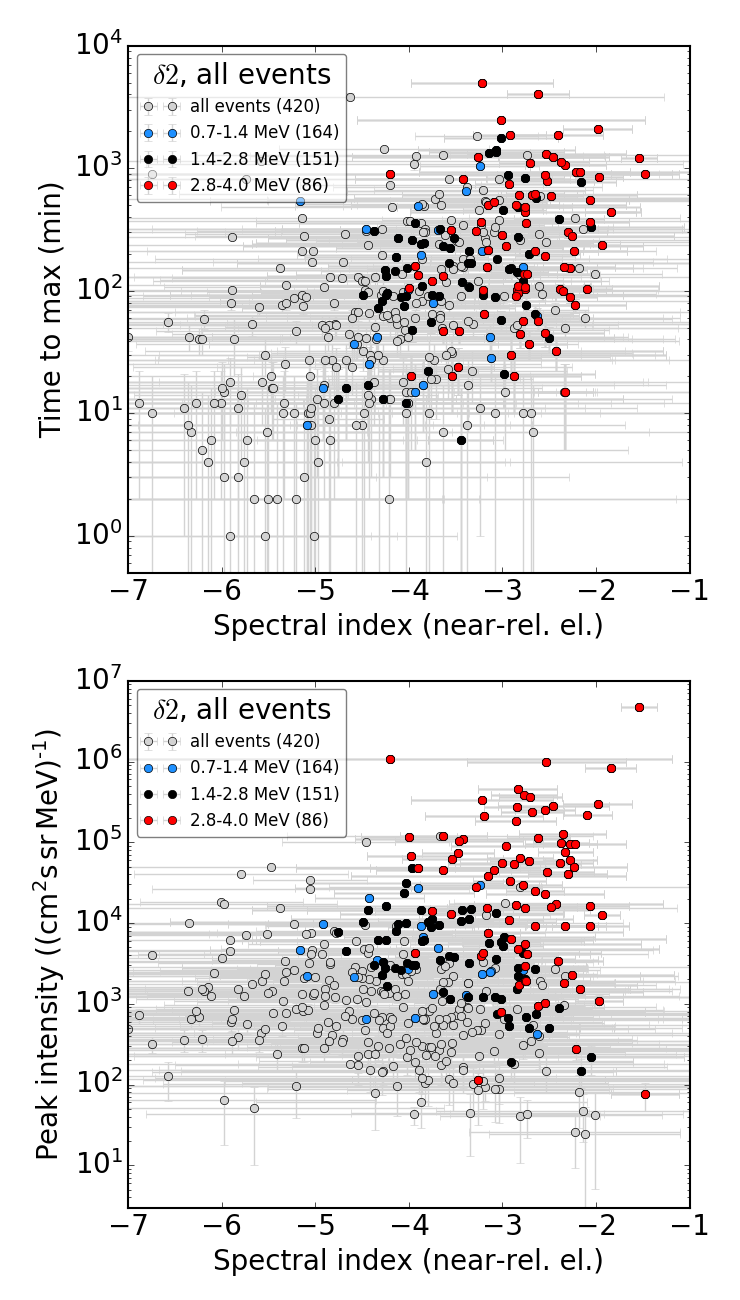}
\caption{Similar to Fig. \ref{fig:gamma_vs_ttm_Imax} but for broken power law events only with $\delta_1$ shown on the left hand side and $\delta_2$ on the right.}
\end{figure*}\label{fig:gamma12_vs_ttm_Imax}
A more detailed view on this dependence is shown in Fig. \ref{fig:ttm_vs_gamma} displaying the rise times as a function of the spectral index at 70\,keV ($\delta70$). 
While the the variation of rise times is very small for events with soft spectral indices ($\delta70\lesssim-4$), being almost exclusively impulsive, a large variation of rise times is observed at the harder-spectra side with ($\delta70\gtrsim-3.5$). 
The separation of events into soft- and hard-spectra events, used in this study, is driven by this difference in the variation of rise times using a limit of $\delta70\sim-3.5$.
Spectra associated with long rise time (gradual) events are almost exclusively hard ($\delta70\gtrsim-3.5$).
Furthermore, there the distribution seems to be limited towards the left hand side, being caused by the absence of soft-spectra events with longer rise times.
Figure \ref{fig:gamma_vs_ttm_Imax} plots the same dependence on a logarithmic scale (to emphasize the short rise time events) in the top panels with $\delta70$ on the left and $\delta200$ on the right. Figure \ref{fig:gamma12_vs_ttm_Imax} shows the same for the broken power law events with $\delta_1$ on the left and $\delta_2$ on the right, respectively. 
The bottom panels show the corresponding distributions of the peak intensities at 55-85\,keV (with pre-event background subtracted) which show the highest values preferentially during hard-spectra (upper right corner). The majority (90\%) of the very high intensity events, with peak intensities $I>1e4$ belong also to the non-impulsive group (not shown).
The lower limit of the peak intensity distributions (bottom panels, lower left corner) is defined by the detection limit of the instrument with the downward slope (from soft to hard spectral indices) being caused by small and soft events being less likely to be detected as the higher energies will then be hidden in the instrumental background. Furthermore, as the event list is based on the energy range of 55-85\,keV, very small and soft events, only significantly observed below that energy, do not appear in the sample. 
The limit at the upper left side of the distribution is not instrumental and is much sharper for $\delta70$ (Fig. \ref{fig:gamma_vs_ttm_Imax} left) than for $\delta200$ (Fig. \ref{fig:gamma_vs_ttm_Imax} right). Its cause is, however, not clear (see section \ref{sec:discussion}).
The colored points in Fig. \ref{fig:gamma_vs_ttm_Imax} denote the presence of MeV electrons as measured by STEREO/HET in its three electron channels. The plotted color (see figure legend) marks the highest HET energy bin, where a corresponding event was observed. The HET events have been identified as significant ($3\sigma$) increases above pre-event background in temporal coincidence with the respective SEPT event. 
These high energy electrons mainly occur during the very intense and hard-spectra events, representing the events with more efficient acceleration, i.e. producing more and higher energy electrons.
23\% of the \ac{nr} electron events are accompanied by 0.7-1.4\,MeV electrons while only 12\% show an increase in the highest HET channel of 2.8-4\,MeV. 
\\
Figures \ref{fig:gamma70_vs_ttm_hign_en_protons} ($\delta70$) and \ref{fig:gamma200_vs_ttm_hign_en_protons} ($\delta200$) separate the events into those with (left) and without (right) an accompanying high-energy proton event as indicated by a coincidental increase (of 5$\sigma$ above pre-event background) in the 60-100\,MeV proton channel of STEREO/HET. 
The fraction of \ac{nr} electron events which are accompanied by 60-100\,MeV protons is 11\%.
The events with accompanying 60-100\,MeV protons (left sides) only populate regions of harder spectral indices.  
Furthermore, most of the highest (55-85\,keV) peak intensities in the sample belong to those events accompanied by 60-100\,MeV protons.
Almost no solely near-relativistic electron event, not detected in the HET electron range (gray points), is observed when the high energy protons are present.
Also most of the MeV electron events (colored points) and especially most of the highest energy electrons (red points) appear on the left hand sides for both ($\delta70$) and ($\delta200$).
On the right hand sides of Figures \ref{fig:gamma70_vs_ttm_hign_en_protons} and \ref{fig:gamma200_vs_ttm_hign_en_protons} (where 60-100\,MeV protons are not present), the MeV electrons accompany the \ac{nr} electrons in only 15\% of the events. 
\begin{figure*}
\plottwo{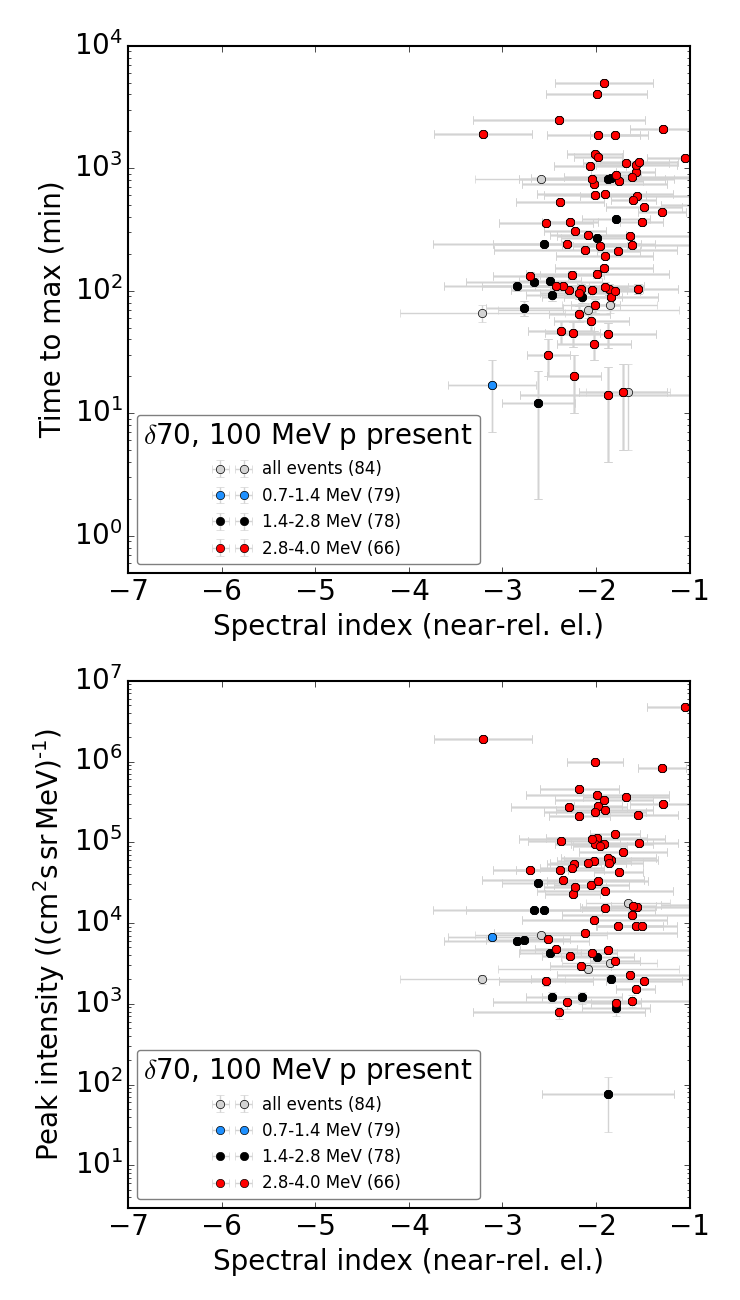}
{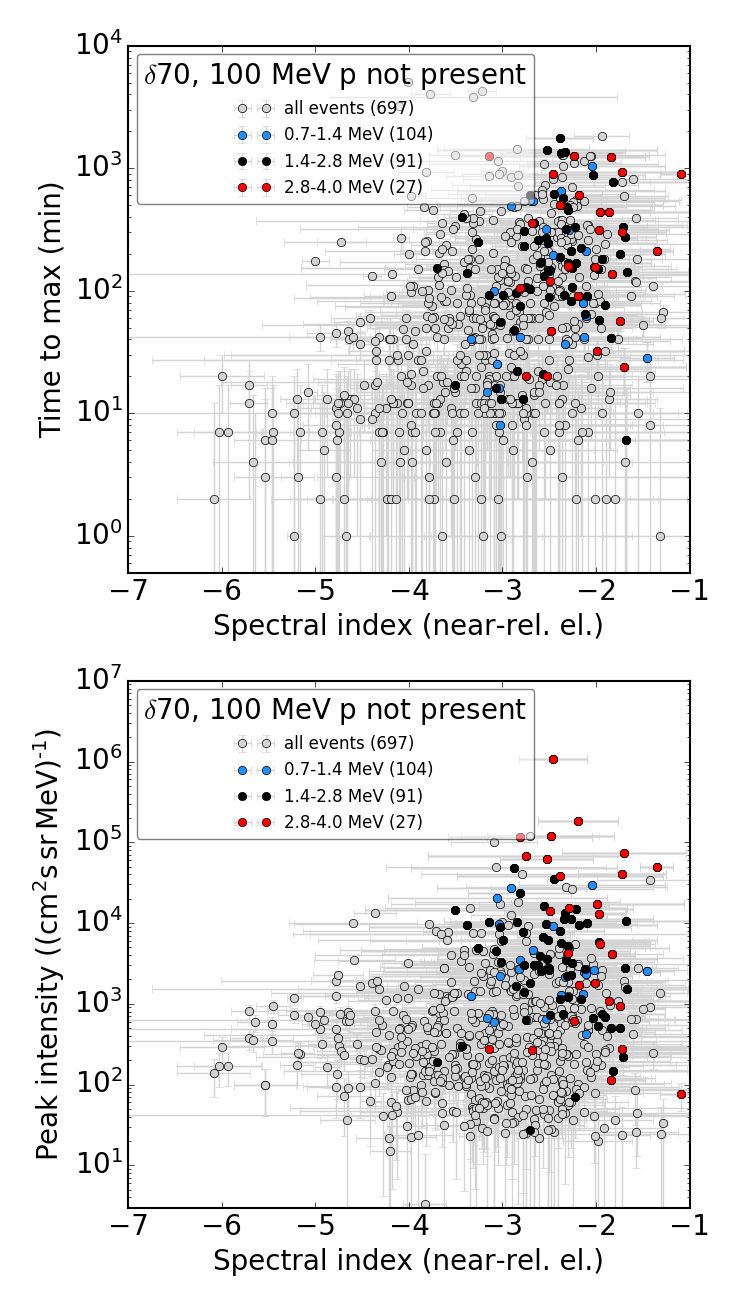}
\caption{Rise times (top) and peak intensities (bottom) at 55-85\,keV as a function of the spectral index at 70\,keV. The left figure shows only events where 60-100\,MeV protons were present, and the right figure where these were not present.}
\end{figure*}\label{fig:gamma70_vs_ttm_hign_en_protons}
\begin{figure*}
\plottwo{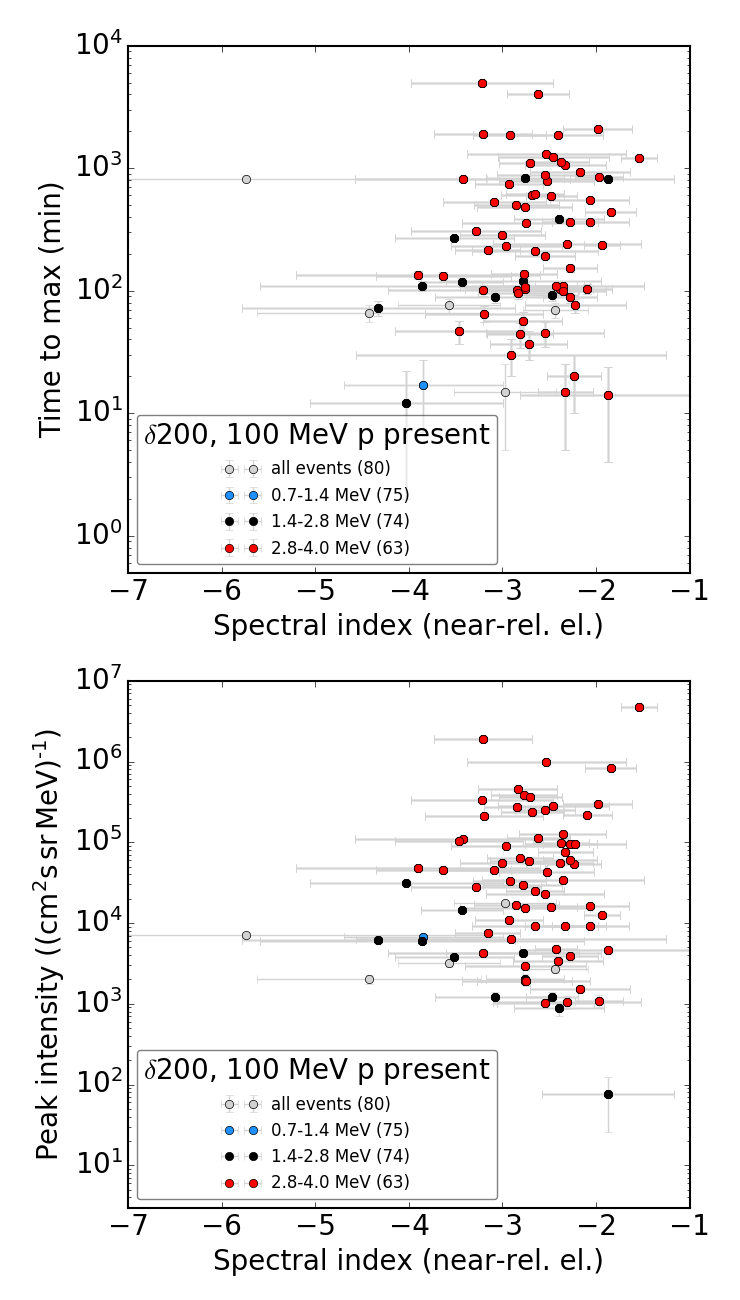}
{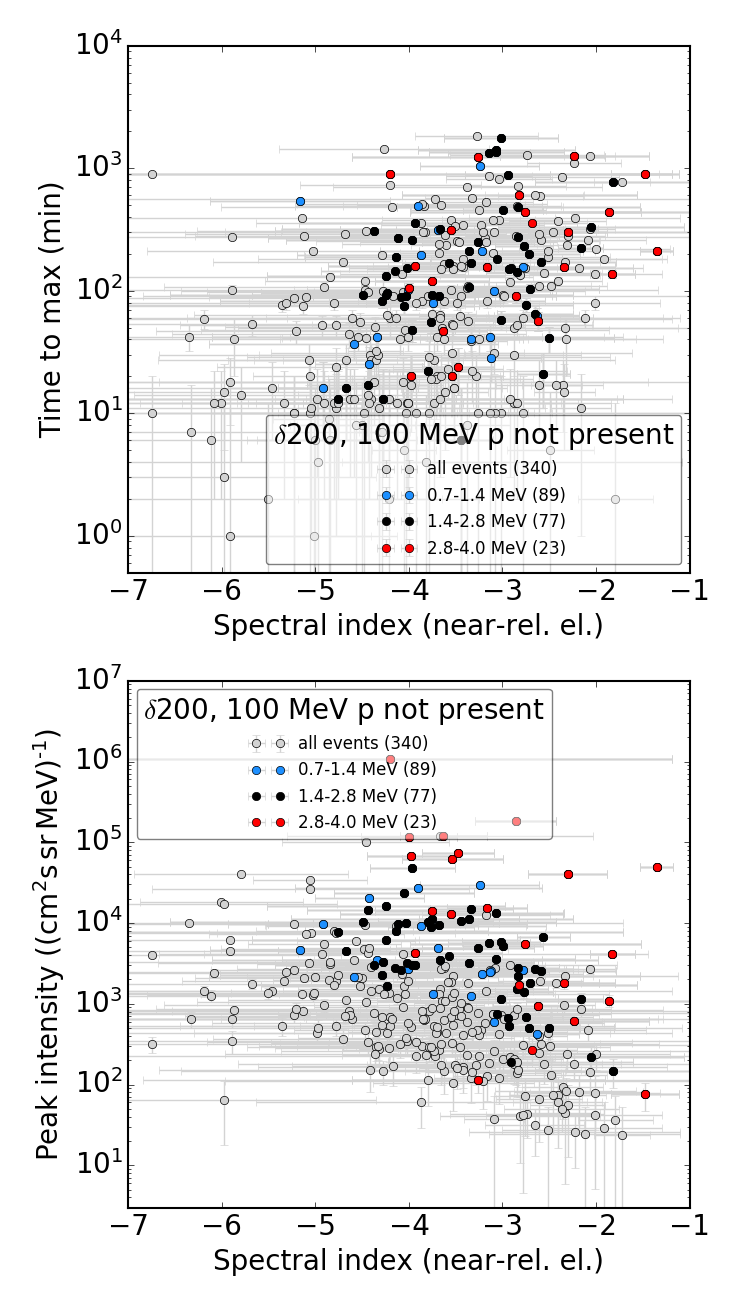}
\caption{Similar to Fig. \ref{fig:gamma70_vs_ttm_hign_en_protons} but for the spectral index at 200\,keV.}
\end{figure*}\label{fig:gamma200_vs_ttm_hign_en_protons}
In the same manner as figures \ref{fig:gamma70_vs_ttm_hign_en_protons} and \ref{fig:gamma200_vs_ttm_hign_en_protons}, figures \ref{fig:gamma70_vs_ttm_typeII} and \ref{fig:gamma200_vs_ttm_typeII} separate the events for cases when a type II radio burst was present (left) and when not (right).
The information on the presence of the type II bursts is taken from the CDAW type II and CME list\footnote{\url{https://cdaw.gsfc.nasa.gov/CME_list/radio/waves_type2.html}}, which is based on Wind/WAVES and STEREO/WAVES data, allowing the type II burst to occur up to 2 hours before the electron onset. 
Because the above list is not comprehensive and contains large gaps, we further complement our list by the type II bursts identified by \cite{Richardson2014} (based on the WIND/WAVES list\footnote{\url{https://solar-radio.gsfc.nasa.gov/wind/data_products.html}}) for $>$25\,MeV proton events if these accompany our electron events. 
Because the source locations haven't been identified for most of the electron events, the 25 MeV proton event list is also used to provide the source locations for a subset of events.
Figures \ref{fig:gamma70_vs_ttm_typeII} and \ref{fig:gamma200_vs_ttm_typeII}, therefore, mainly plot the sub-sample of our electron events, which are also accompanied by $>$25\,MeV protons. The events with type~II radio bursts (left sides of figures \ref{fig:gamma70_vs_ttm_typeII} and \ref{fig:gamma200_vs_ttm_typeII}) 
tend to show the highest peak intensities in the sample. 
A fraction of about 73\% of these events is also accompanied by MeV electrons, while this is only the case for 58\% of the events without type~II (figures \ref{fig:gamma70_vs_ttm_typeII} right). However, there is no striking difference between the distributions with and without type II burst which is most likely due to the bias of the reduced sample of events, where most of the plotted events are accompanied by $>$25\,MeV protons.\\
\begin{figure*}
\plottwo{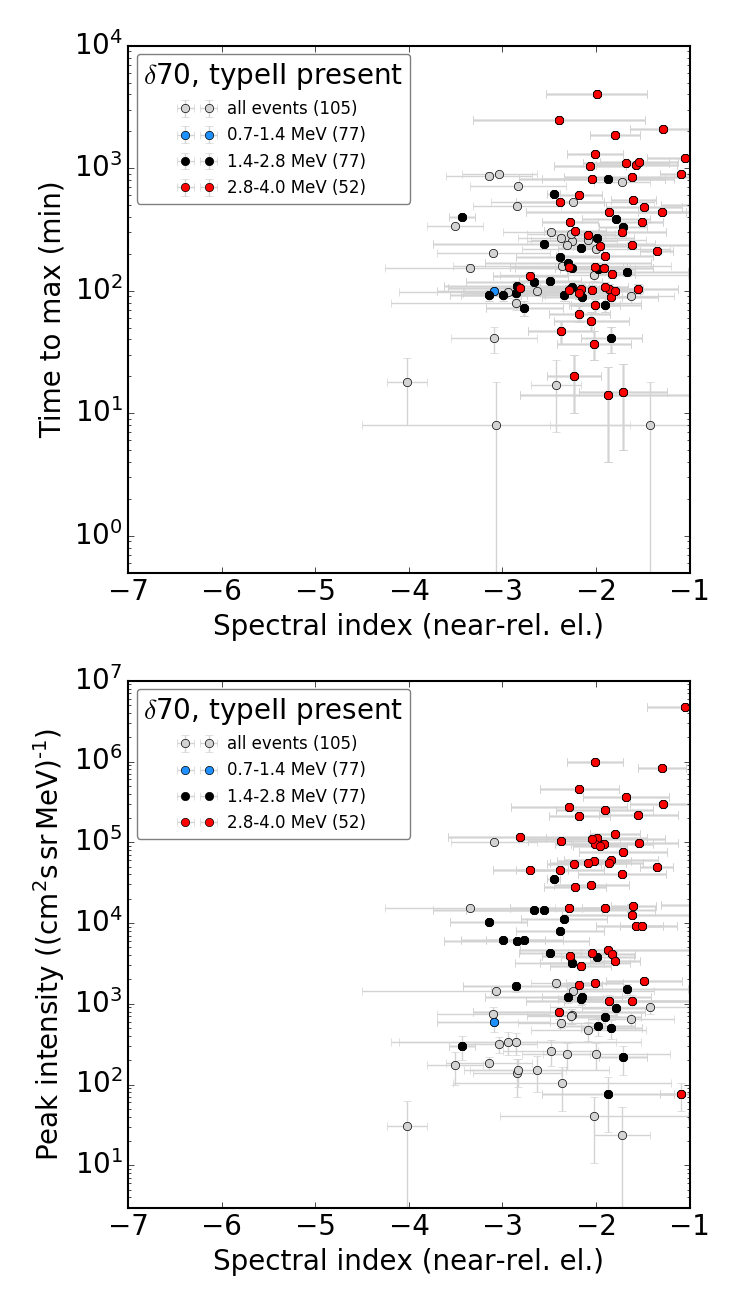}
{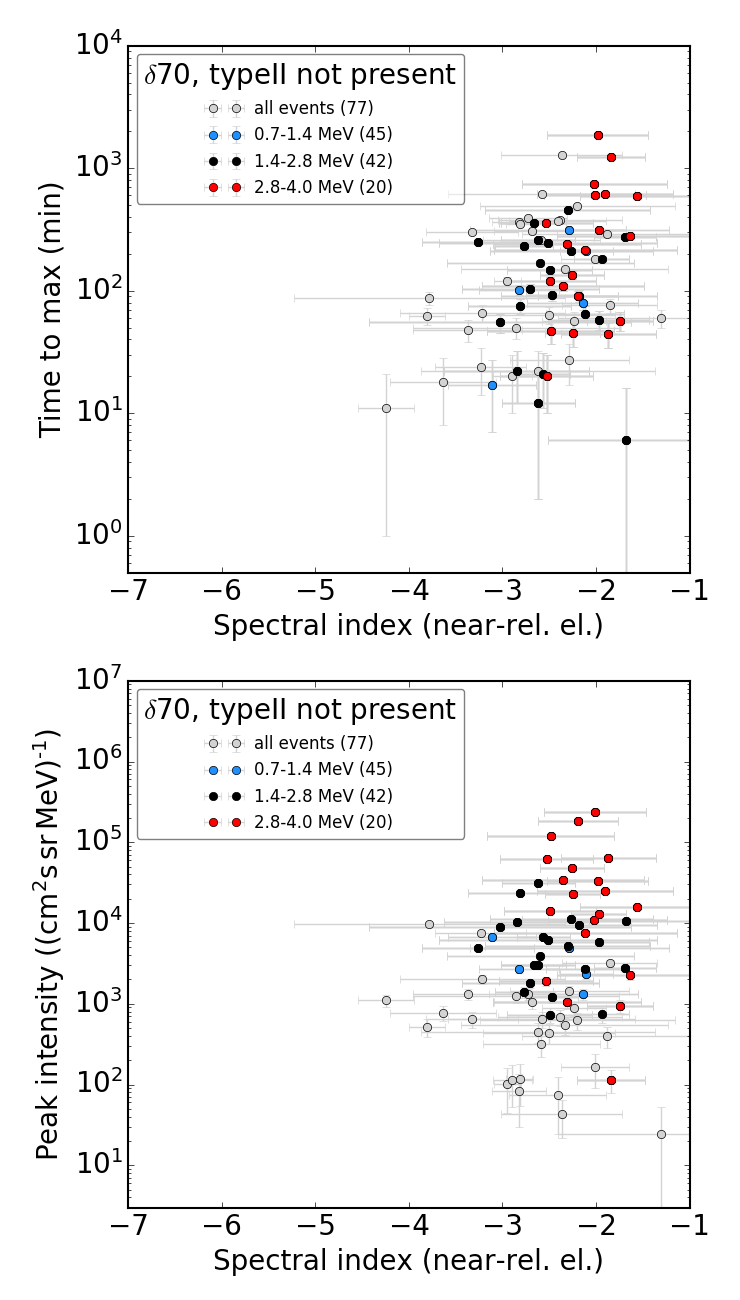}
\caption{Rise times (top) and peak intensities (bottom) at 55-85\,keV as a function of the spectral index at 70\,keV. The left (right) figure shows only events which were (not) accompanied by type~II radio bursts. This plot only includes those events, which were accompanied by $>$25\,MeV protons.}
\end{figure*}\label{fig:gamma70_vs_ttm_typeII}
\begin{figure*}
\plottwo{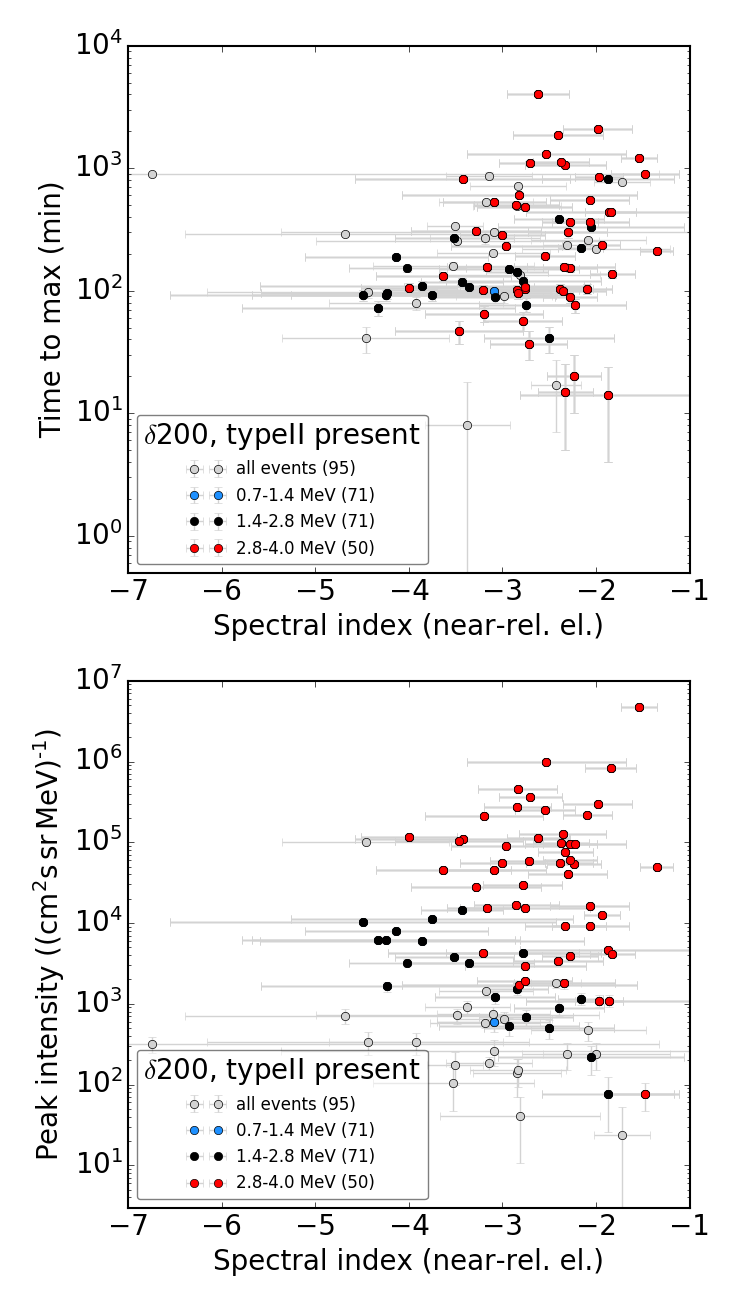}
{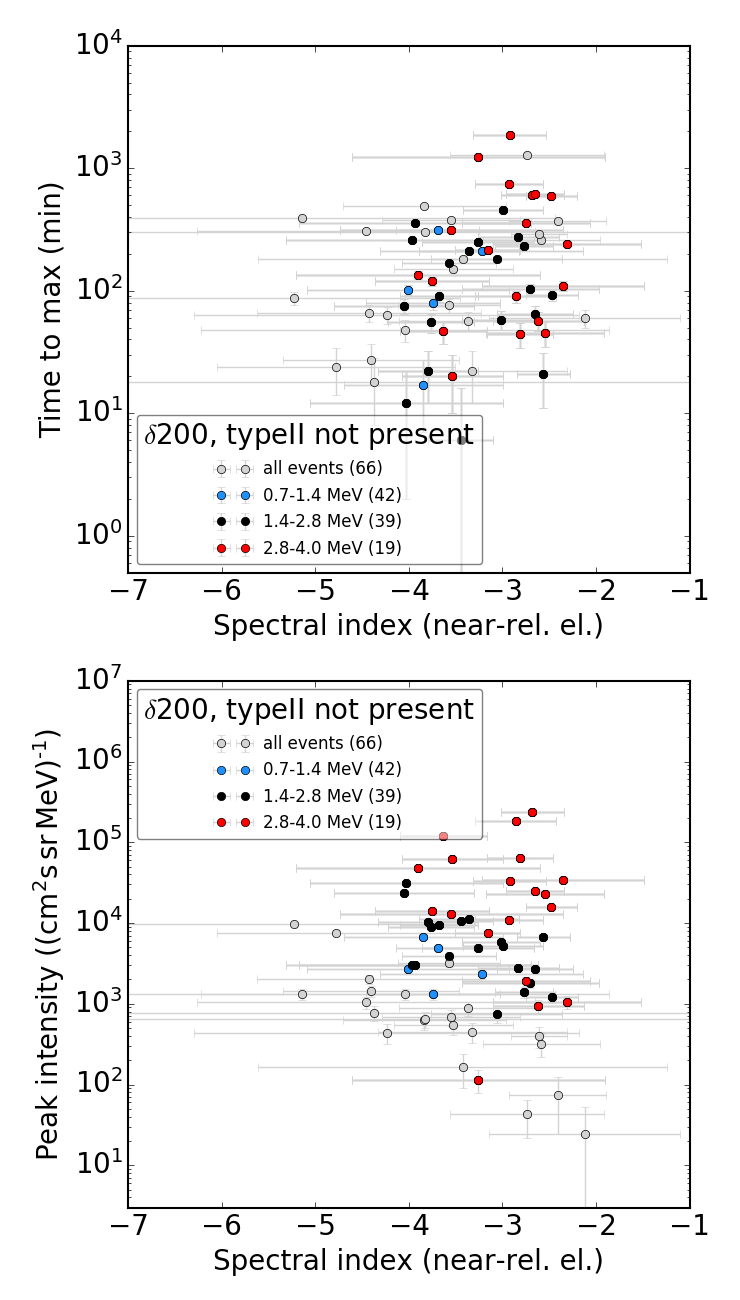}
\caption{Same as Fig. \ref{fig:gamma70_vs_ttm_typeII} but for the spectral index at 200\,keV. This plot only includes those events, which were accompanied by $>$25\,MeV protons.}
\end{figure*}\label{fig:gamma200_vs_ttm_typeII}
For the same reduced sample Fig. \ref{fig:gamma_vs_longsep} plots the spectral indices $\delta70$ and $\delta200$ as functions of the longitudinal separation angle between the parent flaring active region and the magnetic footpoint longitude of the spacecraft. The latter one was determined with a ballistic backmapping to the solar source surface, taking into account the observed solar wind speed. The solar source locations of the events are provided by \cite{Richardson2014}. There is no striking dependence for either $\delta70$ or $\delta200$ on the separation to the parent source region at the Sun. 
However, for $\delta200$ the softest spectra ($\delta\sim-4$) events tend to cluster at well-connected positions (separation angles $\lesssim60^{\circ}$).
\begin{figure*}
\plottwo{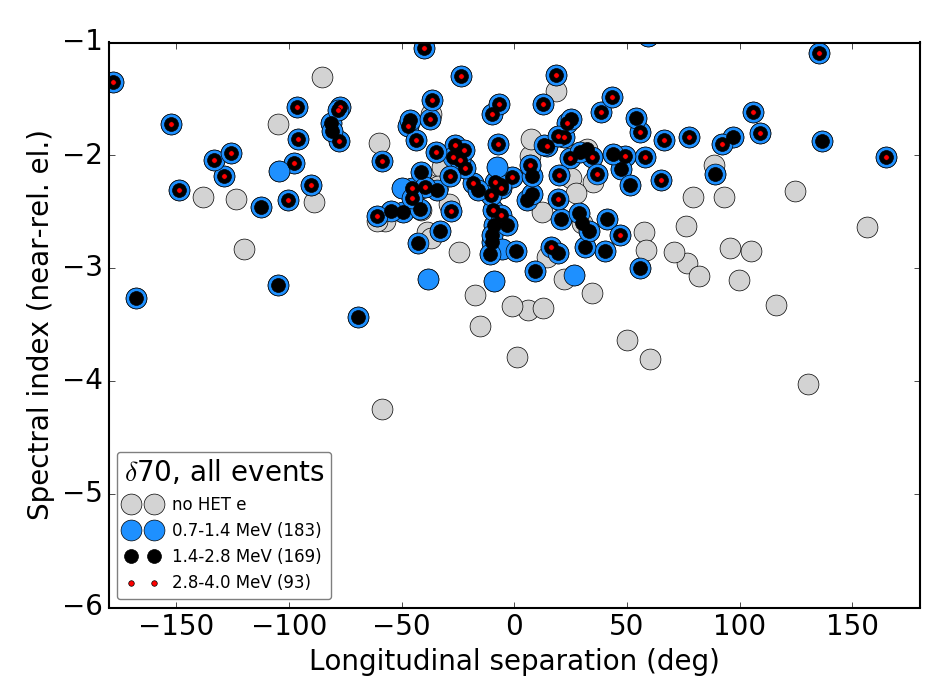}{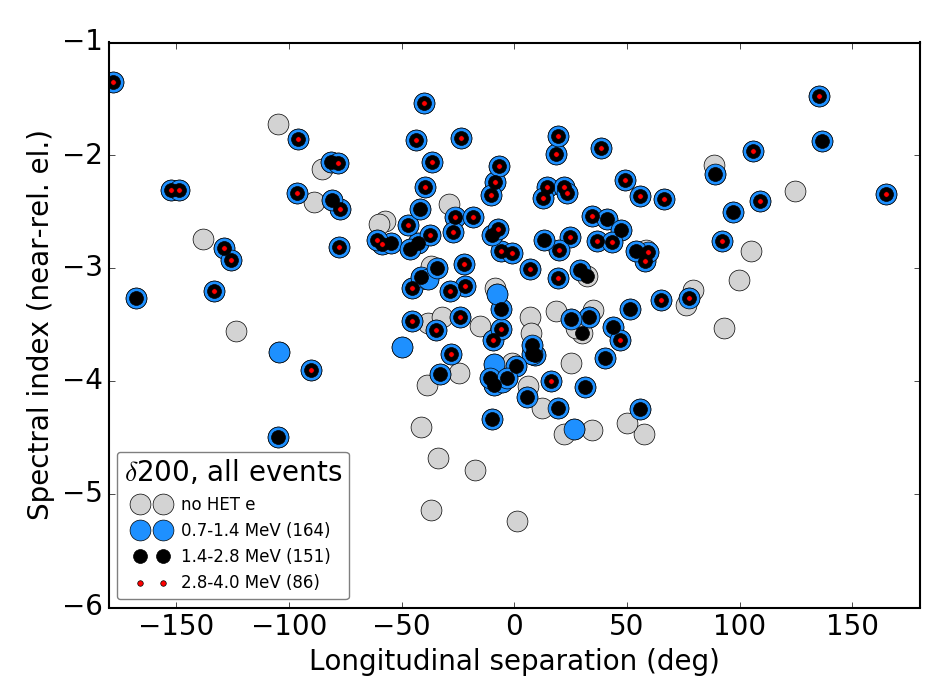}
\caption{Spectral index at 70 and 200\,keV (left and right) as a function of the separation between the flaring region and the backmapped magnetic footpoint longitude of the
spacecraft. This plot only includes those events, which were accompanied by $>$25\,MeV protons. Colored points denote the presence of MeV electrons.}
\end{figure*}\label{fig:gamma_vs_longsep}
%
%
\subsection{Multi-spacecraft Events}
\begin{figure}
\plotone{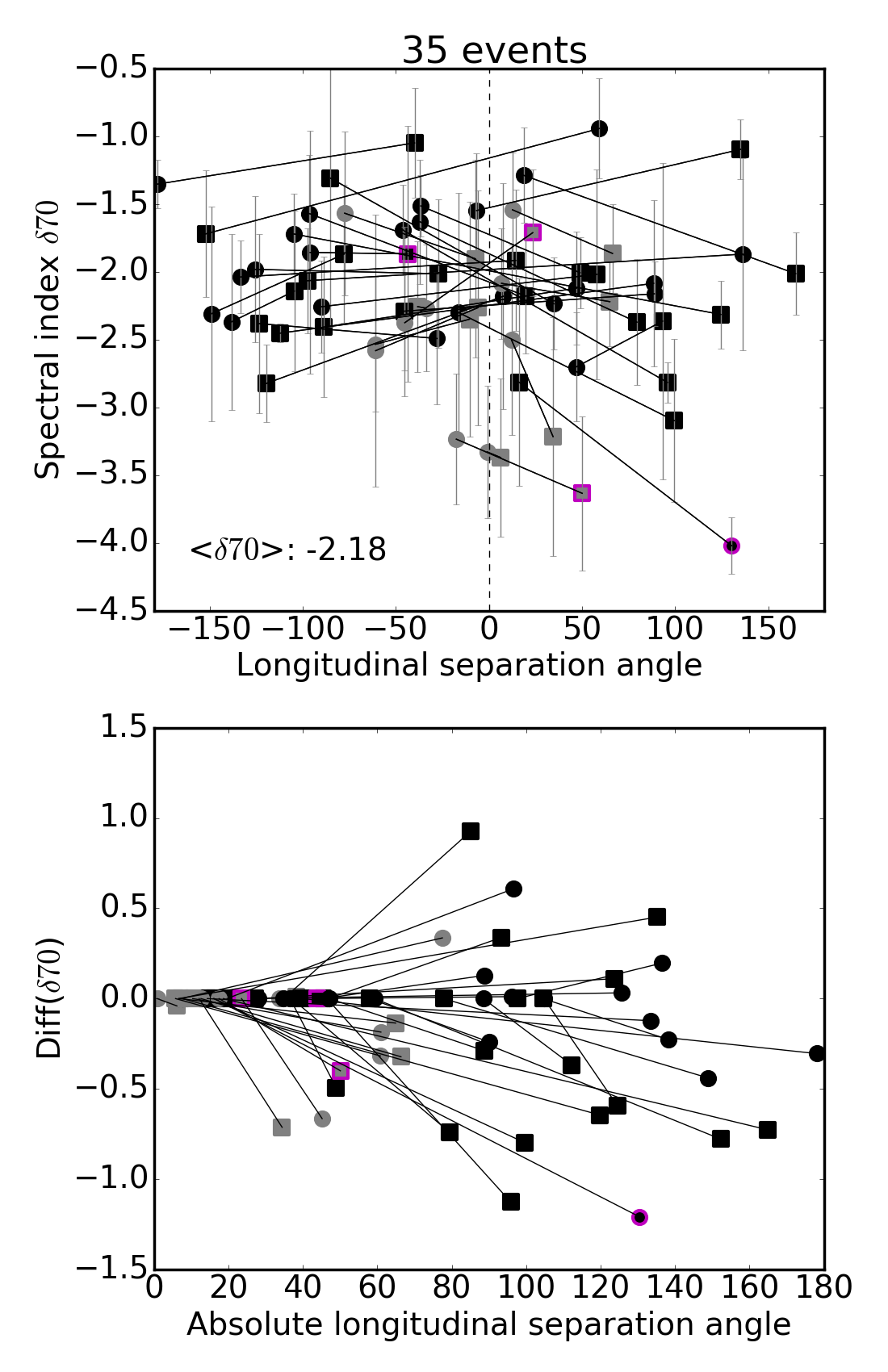}
\caption{Top: Spectral index at 70\,keV as a function of the  separation between the flaring region and the backmapped magnetic footpoint longitude of the spacecraft
for the multi-spacecraft events. Gray (black) points mark narrow spread (widespread) events. Magenta borders mark impulsive events. Bottom: Spectral change from better connected spacecraft to worse-connected one for the same sample of events as a function of the absolute longitudinal separation angle. A negative (positive) change marks a spectral softening (hardening) towards larger separations.}
\end{figure}\label{fig:gamma_multi_sc}
For those events which were accompanied by $>25$\,MeV protons we selected all events observed by both STEREO spacecraft.
Fig. \ref{fig:gamma_multi_sc} (top) shows the spectral indices $\delta70$ of the multi-spacecraft events (points connected by lines) as a function of the longitudinal separation angle. Widespread events, meaning events where the observers are either separated by more than 80 degrees or where one observer is separated more than 80 degrees from the parent flare site, are marked in black while the narrower events are plotted in gray. 
However, every point must be taken as a lower limit as the actual event may be wider than the range covered by the spacecraft.
The bottom panel shows the spectral index change within each event from the better-connected spacecraft to the worse-connected one as a function of the absolute longitudinal separations of each spacecraft to the parent flare.
Most of the events show a spectral softening (for $\delta70$) towards larger separation angles (negative spectral change).
For the spectral index at 200\,keV we do not find such a difference but the numbers of events with softening or hardening are almost equal (not shown). 
The average spectral index for the multi-spacecraft events ($\left<\delta70\right>=-2.18$) is harder than the mean of the whole sample ($\left<\delta70\right>=\mDown$) and the widespread sample shows an even harder mean ($\left<\delta70\right>=-2.07$) than the narrow spread sample ($\left<\delta70\right>=-2.44$).
However, the strong event to event variations, result in widths of the distributions of the spectral index of about 0.7 to 1.1 (as defined like in Table \ref{tab:means}). Because of that together with the low statistics of the multi-spacecraft sample, this apparent spectral hardening must be taken with caution.

\section{Discussion and Conclusions}
\label{sec:discussion}
The presented study draws a comprehensive picture of the spectra of \ac{nr} solar electron events in solar cycle 24 at 1\,au. As the event selection was only driven by the significance level of the electron increases, no bias is present, other than the instrument sensitivity and limitations, which would favour certain types of events. 
For the whole sample of events (\nall) we find hard mean values of the analyzed spectral indices (see Table \ref{tab:means}) with values of $ >\left<\delta\right> > -3$ at lower energies (e.g. $\delta_1$, $\delta70$) and values of $- 7 \left<\delta\right> >-4$ at higher energies (e.g. $\delta_2$, $\delta200$). 
All spectral indices in our sample vary within $-7 < \delta < -1$. 
The mean values soften when only the $\nallImp$ impulsively rising events (with rise times smaller than 20 minutes) are selected. Rise times up to 20 minutes are the most likely ones in the distribution of rise times of our sample (cf. Fig. \ref{fig:hist_ttm}). These impulsive events do, however, only account for about one quarter of all events in the sample.\\
A broken power law spectrum has been reported for impulsive solar electron events \citep[e.g.][]{Lin1982, Lin1985, Krucker2009} and we find such a spectral shape for about half of the events in our sample and for one third of the impulsive group.
It is not clear if the single power law events are really pure power-law spectra or if a potential break point lies outside the detected energy range or outside the range where we are able to detect a break point (70\,keV~$<E_b<300$\,keV) based on the instrument limitations and our fitting procedure.
\\
While \cite{Krucker2009} argued that the origin of the spectral break could either be the acceleration process itself, a secondary process, such as the escape from the acceleration site, or transport effects, \cite{Kontar2009} found that plasma wave generation by electrons below 100\,keV during their propagation from the Sun causes a spectral break. This is because the energy loss of the electrons leads to a flattening/hardening of preferentially the low energy part of the spectrum.
The break point around 35\,keV at 1\,au found by  \cite{Kontar2009} agrees reasonably well with the mean spectral break of $E_b=60$\,keV (ranging from 30 to 100\,keV) found by \cite{Krucker2009}. 
Our study finds a significantly higher break point of about 120\,keV.
Furthermore, a prediction of the energy loss process due to wave generation \citep{Kontar2009} is that the break energy should be correlated with the intensity at the break. However, such a correlation is not observed in the present study (Fig. \ref{fig:break_vs_int_at_break}).
This suggests, that the spectral breaks, found in our study, are not formed by the energy-loss process due to wave generation.
Generally, every energy-loss process, which acts differently on particles of different energies, could lead to the formation of a spectral break.
\cite{Dalla2015} studied the effect of adiabatic cooling of solar energetic protons and noted that particle drifts significantly contribute to this deceleration effect, with larger relative energy losses for lower energy particles. Although the drifts for electrons will be rather small, a significant relative energy loss could still be the result which provides another potential mechanisms for a spectral break.
Additional to energy-loss processes, a particle-loss process could be a further reason for a spectral break. 
Such a particle-loss could be caused by preferential particle scattering at magnetic irregularities off the connecting magnetic field line. The presence of such energy-dependent scattering was found by \cite{Droege2000} and \cite{Agueda2014} through energy dependent transport modeling of \ac{sep} events, showing that the mean free path of \ac{nr} electrons decreases with increasing energy. 
The higher energy electrons therefore experience stronger scattering than the lower energy ones up to a constant mean free path above $\sim1$GV ($\sim1$600\,keV).
For electrons propagating from the Sun to an observer at 1\,au this effect could  result in a depletion of the high energy component because these particles where scattered off (or back along) the connecting field line.
However, it is not clear yet, if the potential mechanisms mentioned above would indeed manifest as a spectral break or only as a systematic change of the single power law.
Other observational studies \citep[e.g.][]{Lin1982, Lin1985} also found break points at higher energies of 100 to 200\,keV.
We therefore suggest that the break point itself and the potential presence of various break points, e.g. one at $\gtrsim100$\,keV and one at $\lesssim60$\,keV, might be caused by different processes or a combination of those as discussed above.
However, one has to keep in mind, that the determined spectral values can be influenced by the fitting range and/or the energy limits and binning of the employed instrument. Our fitting method, applied to \ac{sept} data, limits the break energy to be found between $\sim$70 and $\sim$300\,keV. 
However, the energy binning of \ac{sept} is finer than that of Wind/3DP employed by \cite{Krucker2009}. 
\\
\\
We find a strong variation of rise times for our 55-85\,keV electron events ranging from minutes up to days (see Fig. \ref{fig:hist_ttm}). 
Those events showing short rise times appear consistent with being caused by solar flares, which involve a short-duration (of the order of minutes) acceleration and interplanetary injection.
The events with long rise times, however, need a further process to be involved. On the one hand, a gradual rise can be caused by strong particle diffusion in the interplanetary medium, even if the injection at the Sun is very short. 
However, transport modeling of \ac{nr} solar electron events usually reveals rise times of the order of a few hours, even if the scattering or (perpendicular) diffusion is very strong \citep[cf.][]{Dresing2012, Agueda2014, Droege2014, Droege2016}. Furthermore, these processes will reduce the peak intensity by smearing out the impulsive shape of the distribution on the one hand and by particle loss due to perpendicular diffusion on the other hand.
Since the events with long rise times include some of the largest-intensity events, this also suggests that diffusive transport is not the main process responsible for long rise times. 
Furthermore, diffusive transport should be equally present for all events not depending on their spectral index. So, the absence of longer rise times during soft-spectra events further suggests that the electron diffusion effect is not the most dominant process determining the observed rise times.
Another mechanism responsible for long rise times is a long-lasting acceleration and/or interplanetary injection of the electrons.
The observed correlation of harder spectra (e.g. $\delta70\gtrsim -3.5$) with longer rise times, and the additional presence of higher energy (0.7-4.0\,MeV) electrons for these events (see Fig. \ref{fig:gamma_vs_ttm_Imax}) shows that the very long rise time events are associated with a more efficient acceleration process, yielding higher energies / harder spectra. And this acceleration process must involve a long-lasting electron acceleration or interplanetary injection.\\
The distribution of peak intensities as a function of the spectral index shows a limit towards the upper left with an increasing slope from soft to hard spectral indices (see figures \ref{fig:gamma_vs_ttm_Imax} and \ref{fig:gamma12_vs_ttm_Imax}). This limit is is sharper for the spectral index at lower energies (e.g. $\delta70$ or $\delta1$). The reason for this limit is not clear. A potential explanation would be the effect of energy loss of the lower energy part in the spectrum due to wave generation as discussed above. The higher the electron intensity, the larger the wave-excitation and, therefore, the energy-loss, which would cause stronger spectral hardenings of the low-energy part with increasing peak-intensity. 
However, the missing correlation of the break energy with the intensity (Fig. \ref{fig:break_vs_int_at_break}) suggests, that this effect is not visible in our spectra, probably because \ac{sept} does not cover low enough energies.
\\
The association of our long-rise-time and hard-spectra events with the presence of 60-100\,MeV protons (Fig. \ref{fig:gamma70_vs_ttm_hign_en_protons} and \ref{fig:gamma200_vs_ttm_hign_en_protons}) might suggest a common acceleration process for high energy protons and MeV electrons with the most favourable candidate being the \ac{cme}-driven shock. 
We therefore plot the same distributions distinguishing by the presence or absence of a type II radio burst in figures \ref{fig:gamma70_vs_ttm_typeII} and \ref{fig:gamma200_vs_ttm_typeII}. 
However, here we only take into account those events, where a type II burst was either reported in the (incomplete) CDAW type II list or by \cite{Richardson2014}, meaning that a $>25$\,MeV proton event was also present in the latter case. 
We note, that efficient electron acceleration at shocks, especially to energies $\gtrsim100$\,keV, still challenges state of the art modeling \citep[e.g.][]{Guo2015, Trotta2019} suggesting that the simple presence of a shock may not be enough to explain the electron event.
The very gradual events showing rise times of $\gtrsim 1000$ minutes further challenge the shock acceleration scenario as the shock would have to efficiently accelerate the \ac{nr} electrons over distances of about half an astronomical unit, which is not expected based on in-situ observations of shock-crossings at 1\,au where the shock is very inefficient in accelerating \ac{nr} electrons \citep[e.g.][]{Dresing2016b, Yang2019}.
However, a few of the very long-rise time events ($\gtrsim 1.5$ days) of our sample are accompanied by a \ac{cme} driving a shock where the time scale of the shock propagation to 1\,au roughly fits the electron event rise time. These events might be extreme cases and are subject to future investigations.
Other possible scenarios accounting for the hard-spectra and long-rise time events of our study are ongoing acceleration in post-flare loops \citep{Klein2005}, or re-acceleration of flare particles in the \ac{cme} environment \citep{Petrosian2016} which may not but could involve the presence of a shock.
More complex scenarios involving interacting \acp{cme} \citep{Gopalswamy2004, Li2012}, the presence of magnetic traps \citep{Kocharov2017, Dresing2018} or particle mirrors \citep{Kallenrode2001a, Kallenrode2001b} and consequently enhanced turbulence levels \citep{Xiong2006}, may play a role in understanding how solar energetic electrons are efficiently accelerated and injected into the \ac{ip} medium. 
\\
\\
Finally, we analyzed the spectral indices in terms of a longitudinal dependence.
A spectral systematic spectral change with longitudinal separation from the parent source location could either be caused by transport effects in the \ac{ip} medium or by longitudinal differences of a potentially extended source.
However, whether viewed as single spacecraft (Fig. \ref{fig:gamma_vs_longsep}) or multi-spacecraft (Fig. \ref{fig:gamma_multi_sc}) events, there is no striking dependence of the spectral index on longitudinal separation.
A slight tendency for softer spectra to be observed for well-connected positions is evident in Fig. \ref{fig:gamma_vs_longsep} for the spectral index at 200\,keV, which could be caused by these events being associated with a limited region of weaker acceleration. 
However, our sample is also biased since it only includes events accompanied by $>25$\,MeV protons for which the sources were identified by \cite{Richardson2014}.
The same is true for the multi-spacecraft events (Fig. \ref{fig:gamma_multi_sc}). No systematic spectral change is seen within the same event as function of the longitudinal separation. 

\section{Summary}\label{sec:summary}
The presented study is a comprehensive analysis of \ac{nr} electron spectra in the energy range of 45-425\,keV observed at 1\,au during solar cycle 24 from 2007 to 2018 by the STEREO mission. 
29\% of these events show impulsive rise times to peak intensity of $<$20 minutes), while the majority show longer, non-impulsive, rise times.
The events are approximately equally divided between those having single power law or broken power law energy spectra.
However, about half of the single power law events do not extend to high enough energies to exclude the potential presence of a break point.
The mean spectral indices in the lower energy range around 70\,keV and in a higher energy regime around 200\,keV are $\delta70=\mDown$ and $\delta200=\mUp$, respectively. 
The spectra are softer for the impulsive group (with rise times $<20$ min) with $\delta70_i=\mDownImp$ and $\delta200_i=\mUpImp$. 
For those events showing a broken power law we find a mean break energy of $E_b=123$\,keV. We do not find a dependence of the break energy on the intensity suggesting that this break is not caused by energy-loss of the lower energy part due to interaction with electrostatic waves in the plasma.
Were the spectral index dependent on this effect, higher peak  intensities should yield higher spectral breaks. This is because a higher electron beam density would cause a faster generation of waves and therefore a stronger interaction between electrons and plasma waves. \citep{Kontar2009}. 
An analysis of the spectral indices and the rise times of the events reveals:
\begin{itemize}
    \item A strong variation of event rise times (from onset to peak intensity) is found ranging from minutes to several days.
    \item Soft-spectra ($\delta\lesssim-4$) events show almost always impulsive rise times but the vice versa is not true as the presence of impulsive events with harder spectra shows. 
    \item Long-rise-time events are associated with harder spectra while the vice versa is not true (there are also impulsive events with harder spectra).
    \item The presence of MeV electrons is associated with the hard-spectra and long-rise-time events
\end{itemize}
While it is likely that the impulsive, soft-spectra events are flare-related events, the hard-spectra, long-rise time events cannot be explained by a simple flare scenario but require another or a secondary process which involves a prolonged particle injection and leads to more efficient electron acceleration (in terms of energy). We note, however, that a potential CME- or shock-related source cannot be excluded for the impulsive events. However, such potential shock-acceleration would have to occur on correspondingly shorter time scales.
\\
\\
We furthermore find a close correlation between the presence of 60-100\,MeV protons with those of the MeV electrons: There are almost no \ac{nr} electron events that are accompanied by 60-100\,MeV protons but not also associated with MeV electrons. This result might imply a common acceleration process of the high energy electrons and protons or a common ingredient for the acceleration processes. 
On the other hand, there are a large number of \ac{nr} electron events not associated with 60-100\,MeV protons that still extend into the MeV range.\\
For a subgroup of events which are accompanied by $>25$\,MeV protons, we find
\begin{itemize}    
    \item no clear dependence of the spectral index on the presence or absence of a type II burst,
    \item no clear dependence of the electron spectral index on the longitudinal separation angle between the spacecraft magnetic footpoint and the parent solar source region, 
    \item no clear systematic dependence of the spectral index on the longitudinal separation angle for multi-spacecraft events.
\end{itemize}
A future analysis involving additional observations close to Earth or the Sun (e.g. from Parker Solar Probe and Solar Orbiter) may shed more light on the possible longitudinal and radial dependence of the spectral index during solar energetic electron events.


\acknowledgments
The STEREO/SEPT, Chandra/EPHIN and SOHO/EPHIN project is supported under grant 50OC1702 by the Federal Ministry of Economics and Technology on the basis of a decision by the German Bundestag.\\
RGH acknowledges the Spanish MICIU/FEDER/AEI for the financial support under project ESP2017-88436-R. 
IGR acknowledges support from NASA programs  NNH17ZDA001N and NH14ZDA001N-LWS, and from the STEREO mission.
FE and ND acknowledge support from NASA grant NNX17AK25G. Additional support from an Alexander von Humboldt group linkage program is appreciated. We thank the International Space Science Institute (ISSI) for hosting our team on 'Solar flare acceleration signatures and their connection to solar energetic particles'. 
Financial support by the Deutsche Forschungsgemeinschaft (DFG) via project  GI 1352/1-1 is acknowledged.\\
We would like to thank Dr.~R.~D. Strauss, Dr.~E.~Kontar, and Dr.~A.~Warmuth for helpful discussions.

\bibliography{references.bib}

\end{document}